\documentclass[pra,aps,groupedaddress,10pt,floatfix,twocolumn,amsmath,amssymb,longbibliography,notitlepage,superscriptaddress]{revtex4-1}

\usepackage{graphicx}
\usepackage{dcolumn}
\usepackage{bm}
\usepackage[pdftex,bookmarks=true,bookmarksnumbered=true,plainpages=false,pdfpagelabels]{hyperref}
\usepackage{units}
\usepackage{url}
\usepackage[T1]{fontenc}
\usepackage[pdftex,usenames,dvipsnames]{color} 
\usepackage{lipsum}
\usepackage{ulem}

\usepackage{pdfsync}

\usepackage[left]{lineno}

\newcommand{\abs}[1]{\left\vert#1\right\vert}

\newcommand{\expect}[1]{\left\langle#1\right\rangle}

\usepackage{physics}

\definecolor{red}{rgb}{1.00,0.00,0.00}
\definecolor{blue}{rgb}{0.00,0.00,0.50}
\definecolor{green}{rgb}{0.99,0.09,0.50} 

\begin{document}

\title{Improving broadband displacement detection with quantum correlations}
\author{N. S. Kampel}
\email{nir.kampel@jila.colorado.edu}
\author{R. W. Peterson}
\author{R. Fischer}
\author{P.-L. Yu}
\altaffiliation[Current address: ]{School of Electrical and Computer Engineering, Birck Nanotechnology Center, Purdue University, West Lafayette, IN 47907 USA}
\affiliation{JILA, University of Colorado and National Institute of Standards and Technology,and Department of Physics, University of Colorado, Boulder, Colorado 80309, USA}
\author{K. Cicak}
\affiliation{National Institute of Standards and Technology, Boulder, Colorado 80305, USA}
\author{R. W. Simmonds}
\affiliation{National Institute of Standards and Technology, Boulder, Colorado 80305, USA}
\author{K. W. Lehnert}
\author{C. A. Regal}
\email{regal@jila.colorado.edu}
\affiliation{JILA, University of Colorado and National Institute of Standards and Technology,and Department of Physics, University of Colorado, Boulder, Colorado 80309, USA}

\date{\today}

\begin{abstract}
Interferometers enable ultrasensitive measurement in a wide array of applications from gravitational wave searches to force microscopes.  The role of quantum mechanics in the metrological limits of interferometers has a rich history, and a large number of techniques to surpass conventional limits have been proposed.  In a typical measurement configuration, the tradeoff between the probe's shot noise (imprecision) and its quantum backaction results in what is known as the standard quantum limit (SQL).  In this work we investigate how quantum correlations accessed by modifying the readout of the interferometer can access physics beyond the SQL and improve displacement sensitivity.  Specifically, we use an optical cavity to probe the motion of a silicon nitride membrane off mechanical resonance, as one would do in a broadband displacement or force measurement, and observe sensitivity better than the SQL dictates for our quantum efficiency.  Our measurement illustrates the core idea behind a technique known as \textit{variational readout}, in which the optical readout quadrature is changed as a function of frequency to improve broadband displacement detection.  And more generally our result is a salient example of how correlations can aid sensing in the presence of backaction.
\end{abstract}

\maketitle

\begin{figure*}[]
	\centering
    \includegraphics[trim={1.25cm 15.25cm 0cm 0cm},clip]{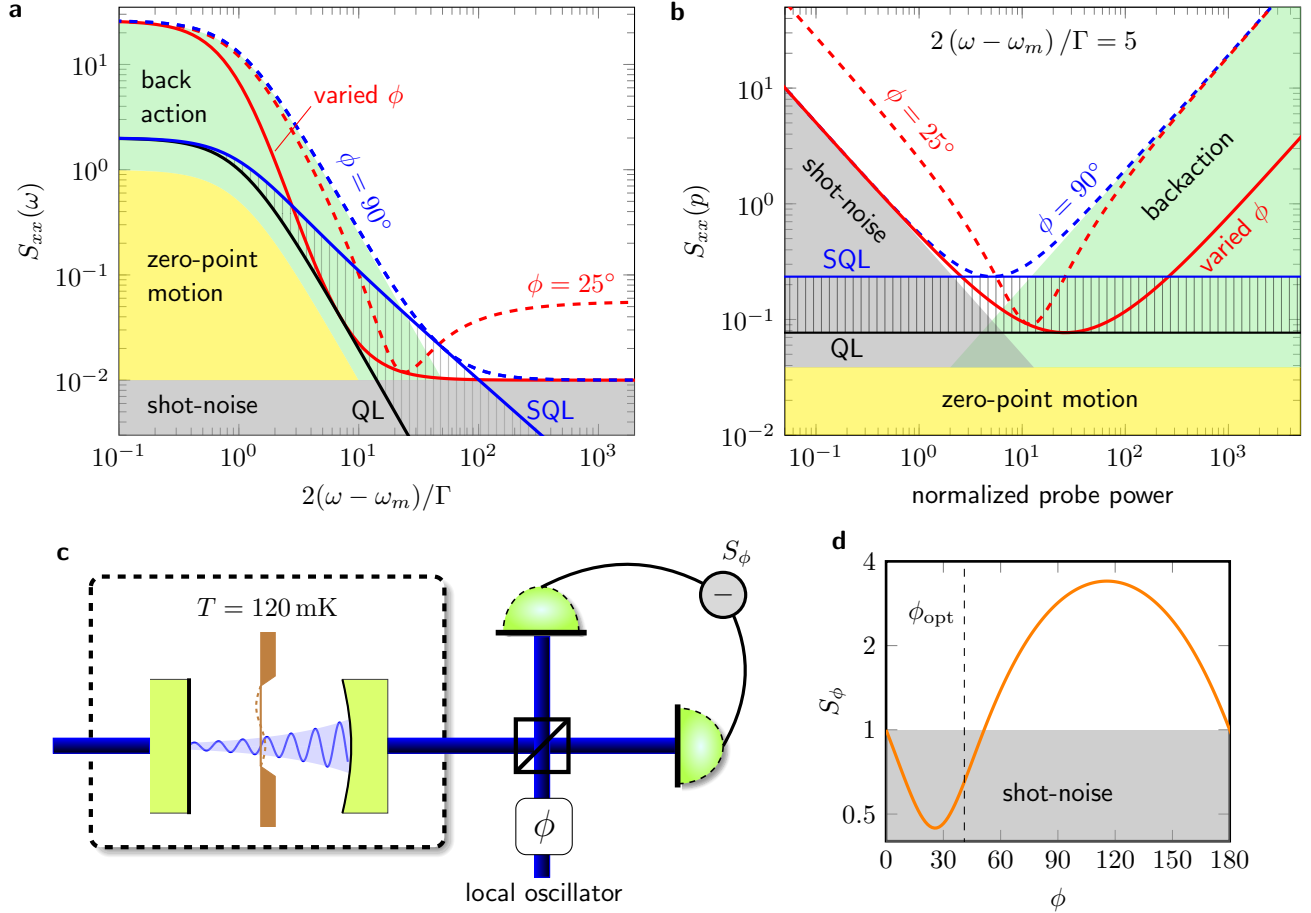}   
	\caption{Imprecision for different interferometric measurement configurations. (a) Frequency dependence for fixed power $p=50$ and (b) Power dependence for frequency $2(\omega-\omega_m)/\Gamma=5$.  Shaded areas represent the SN (gray), backaction (green) and zero-point motion (yellow). Lines represent measurement at $\phi=90^{\circ}$ (dashed blue), $\phi=25^{\circ}$ (dashed red), variational readout (solid red).  The black (QL) and blue (SQL) lines in both (a) and (b) show total limits when the probe power is allowed to vary to optimize total noise, with hatched area highlighting the QL and SQL distinction.  All curves are evaluated for ideal quantum efficiency $\epsilon=1$, and zero thermal disturbance. (c) Experimental schematic. Coherent probe beam enters the cavity and interacts with a membrane resonator. The probe light that leaves the cavity is detected via balanced homodyne detection with the measurement angle (quadrature) set by the local oscillator (LO) phase $\phi$. (d) Comparison to ponderomotive squeezing for frequency $2(\omega-\omega_m)/\Gamma = 5$ and $p=6$. Plotted is the spectral density for the optical output $S_{\phi}$ (orange line) compared to the shot-noise limit (shaded gray) (see Appendix). The dashed black line demonstrates that the optimal mechanical displacement measuring angle ($\phi_{\rm{opt}}$) is shifted towards the phase quadrature compared to the phase for maximal ponderomotive squeezing. }
	\label{fig: exp setup}
\end{figure*}

When one seeks knowledge of the full dynamics of the displacement of a harmonic oscillator, non-commutation of the two mechanical quadratures requires a minimum added noise equal to the mechanical resonator's zero point motion~\cite{Haus62,Caves82}.  This fundamental quantum limit (QL) is a distinct bound from the standard quantum limit (SQL) that is often considered in interferometric displacement measurement~\cite{Braginsky(03)_NonRelaventSQL,Clerk(10)_QN,Jaekel(90)_QL}.  The SQL is a consequence of the non-commutation of the probe's quadratures in a specific measurement configuration and is characterized by a tradeoff between shot noise (SN) imprecision and quantum backaction that are uncorrelated~\cite{Braginskii(68)_SQLfirst,Caves(80)_QM_RPinInter}.  The QL and SQL reach the same limit when probing at the peak mechanical response, but the QL can be a significantly lower bound off resonance  [Fig.~\ref{fig: exp setup}(a,b)].  In studies of micro-mechanical motion, there has been great interest in observing quantum backaction and approaching the SQL on mechanical resonance~\cite{LaHaye(04)_appQL,Purdy(12)_RPSN,SafaviNaeini(12)_RPSN,Schreppler(14)_SQL,Teufel(16)_StrongRPSN}.  However, much of what historically motivates SQL research is displacement monitoring over a wide frequency band, such as gravitational wave searches~\cite{Miao(14)_Interef_lim_GW}.  One long-standing concept for surpassing the SQL is to introduce correlations by changing the readout quadrature as a function of frequency in a technique known as variational readout~\cite{Vyatchanin,Kimble(01)_GI_inter}.  In the work presented in this article, we measure the displacement of a membrane resonator in an optical interferometer with a tunable readout quadrature.  By thermalizing the mechanical device to a dilution refrigerator and mitigating other technical noise sources, we are able to measure quantum noise far off resonance compared to typical micromechanical measurements, and achieve near-SQL-limited measurement of a solid-state object.  With this starting point we are able to use variational-readout techniques to improve upon the off-resonance SQL for our quantum efficiency.

Over the years a variety of techniques have been considered for surpassing the SQL~\cite{Braginskii(68)_SQLfirst}, and it is useful to place these in context in comparison to variational readout.  Perhaps the most well-known technique is to restrict knowledge to a dynamically-decoupled single mechanical quadrature in a quantum non-demolition (QND) measurement to evade backaction completely~\cite{Braginsky(80)_QNDdef}.  Therefore the total broadband noise can be arbitrarily decreased (up to the zero-point motion) by increasing the probe power.  However, QND unfortunately measures only a single phase force, unless the system dimensionality is increased to perform QND measurements on both quadratures~\cite{Hammerer(09)_EPRchannelsMechAtom,Caves(12)_EvadQMmanyQuad,Ockeloen-Korppia2016,Moller2016}. While QND measurements have been demonstrated electromechanicaly~\cite{Suh(14)_MechBAE,Wollman(15)_MechSqueezing,Lecocq(15)_MechSqueezing,Pirkkalainen(15)_MechSqueezing}, instabilities can arise, and a measurement below the SQL has not been demonstrated.  In our measurements, we focus on a solution in which a two-mechanical quadrature measurement is made, yet quantum correlations between imprecision and backaction are used to address the SQL.  Such correlations can be achieved by injecting quadrature squeezed light~\cite{Kimble(01)_GI_inter} or using nonlinear cavities~\cite{Rehbein2005,Laflamme2011}, but an equally-capable technique is to rotate the readout quadrature such that the mechanical motion itself mixes the probe's amplitude and phase quadratures~\cite{BraginskyKhalili(92)_book,Kimble(01)_GI_inter}.

For any given frequency, there exists a measurement configuration that can reach the QL for an ideal detector (black line in Fig.~\ref{fig: exp setup}(a)).  Here we present analysis based upon the probe's uncertainty relations that reveals this configuration~\cite{Robertson(34)_statUncePrinc,Braginsky(03)_NonRelaventSQL,Clerk(10)_QN}.  (Note the on-resonance SQL can be conveniently arrived at with a Heisenberg microscope argument that dictates a minimum contribution of the SN and backaction~\cite{Heisenberg(27)_uc}, but this argument does not hold in the presence of correlation.) We consider the result of a linear measurement of a mechanical harmonic oscillator via an optical probe at a given quadrature angle $\phi$. The analysis is based upon an optomechanical interaction in which probe's phase quadrature linearly depends on the mechanical state, while the amplitude is unchanged.  In the analysis of this interaction we assume a large-photon limit in which the fluctuations are linearized around a large optical field.  From the optical probe one can infer a mechanical displacement, and we present our data and theoretical comparison in terms of a dimensionless mechanical displacement.  This transformation for an optomechanical interaction in the presence of finite efficiency and conversion to dimensionless units is nontrivial, and the expressions put forth here are derived sequentially in the Appendix to enable concise analysis of their consequences on measurements.

The dimensionless mechanical displacement inferred from the optical probe can be written as:  $\hat{X}_{\phi}(\omega) = \hat{x}_{m}(\omega) + \hat{I}_{\phi} - i \tilde{\chi}_m(\omega) \hat{F}_{\rm{ba}}$.   Here the mechanical state ($\hat{x}_{m}$) appears along with SN ($\hat{I}_{\phi}$) and backaction ($\hat{F}_{\rm{ba}}$), where the SN and backaction both arise from the probe's fluctuations.  The SN is a white noise with a contribution that depends on $\phi$.  The backaction term is a result of probe amplitude fluctuations driving the mechanical state and hence is function only of the amplitude quadrature (AM, $\phi=0^{\circ}$).  The backaction is filtered by $\tilde{\chi}_m(\omega)$, the dimensionless mechanical susceptibility given by $\tilde{\chi}_m(\omega)=(1-2i(\omega-\omega_m)/\Gamma)^{-1}$ where $\Gamma$ is the effective mechanical linewidth.  Within $\hat{x}_{m}$ we include both the zero-point motion and environmental perturbation (thermal and other applied forces).

In the experiment we will measure the symmetrized displacement power spectral density (PSD) $S_{xx}(\omega) = \expect{\hat{X}_{\phi}(-\omega)\hat{X}_{\phi}(\omega)}$~\cite{Clerk(10)_QN}, which has contributions from the mechanical resonator ($S_m$), SN imprecision ($S_{II}$), backaction ($S_{FF}$), and their cross-correlation ($S_{IF}$)
\begin{align}
S_{xx}(\omega,p,\phi) = &S_{m}(\omega) + S_{II}(p,\phi) + \abs{\tilde{\chi}_m(\omega)}^2 S_{FF}(p) + \nonumber \\
& 2 \Im[\tilde{\chi}_m(\omega) S_{IF}(\phi)]. \label{eq: pos spec}
\end{align}
Throughout we use dimensionless displacement units such that the zero-point motion contribution to the PSD $S_{xx}$ is $1$ on mechanical resonance ($\omega=\omega_m$), and the probe power ($p$) is normalized to the SQL power on mechanical resonance.  Similarly the value of the added noise at the SQL (SN and on-resonance backaction) is equal to $1$.  In absolute units the added noise at the SQL and the zero-point motion each contribute $S_{\rm{sql}}(\omega_m)= 2 x_{\rm{zp}}^2 /\Gamma$.  The zero-point motion is given by $x_{\rm{zp}}=\sqrt{\frac{\hbar}{2m\omega_m}}$, where $\omega_m$ is the resonant frequency of the mechanical mode of interest, $\hbar$ is the reduced Planck constant, and $m$ is the effective mass of the resonator~\cite{Purdy(12)_RPSN}.

From an analysis of the full expression of $S_{xx}$ shown in the Appendix one finds the probe's uncertainty relation~\cite{Robertson(34)_statUncePrinc} is connected to the measured PSD via  $S_{II} S_{FF} \ge \frac{1}{4} + S_{IF}^2$. The SQL corresponds to the special case in which $S_{IF}=0$, and we will find the QL by careful choice of $S_{IF}$ and measurement power for a given frequency \cite{Clerk(10)_QN,Sudhir(16)_corr}.  To understand measurement limits, we plot the contributions to $S_{xx}$ in Fig.~\ref{fig: exp setup}.  First, we describe the lines relevant to SQL physics.  The dashed blue line in Fig.~\ref{fig: exp setup} shows a phase quadrature (PM, $\phi=90^{\circ}$) measurement for a fixed backaction-dominated probe power, whereas the blue solid line (SQL) results when optimizing the probe power at each frequency.  Due to the backaction frequency dependence, the SQL value changes off resonance as $S_{\rm{sql}}(\omega) = S_{\rm{sql}}(\omega_m)\abs{\tilde{\chi}_m(\omega)}$; namely the SQL value drops and the power required to reach it increases. Figure~\ref{fig: exp setup}(b) shows the SQL results in a linear tradeoff between SN and backaction~\cite{Caves(80)_QM_RPinInter}.

\begin{figure*}[t]
	\centering
    \includegraphics[width=\textwidth,trim={1.75cm 20.5cm 2.75cm 0cm},clip]{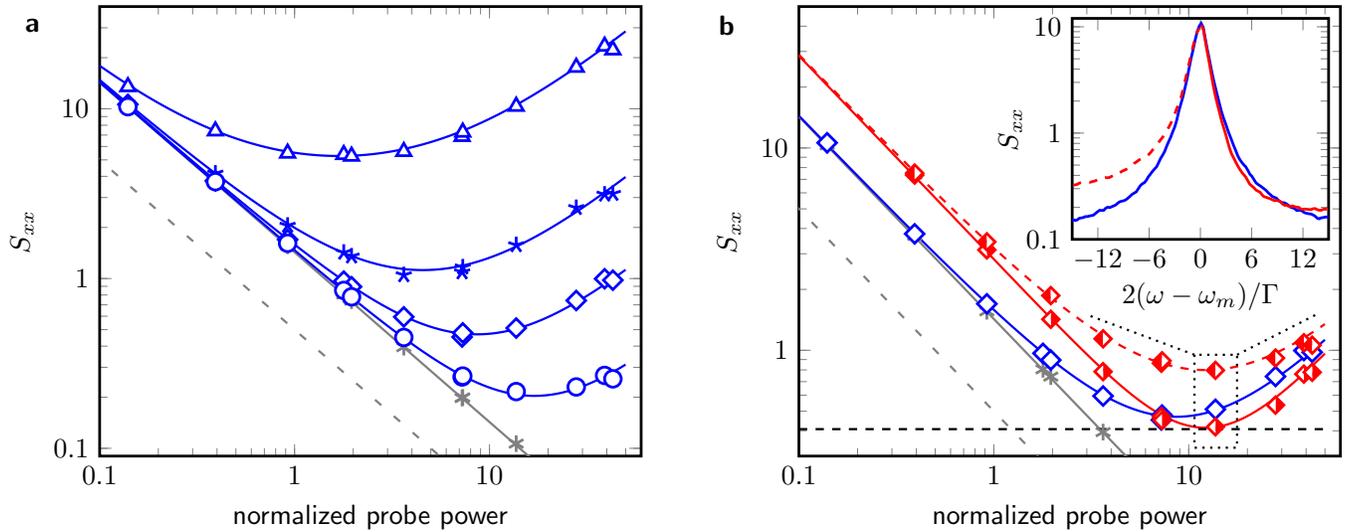} 
	\caption{Measured displacement PSD for different frequencies and measurement angles as a function of normalized probe power p. (a) Measurement at $\phi=90^{\circ}$ at different frequencies: $2(\omega-\omega_m)/\Gamma=0$ (triangle), $2.5$ (star), $5$ (square), $10$ (circle).  (b) Measurement at $\phi=90^{\circ}$ (blue) and at $\phi=45^{\circ}$ (red) compared to SN level at $\phi=90^{\circ}$ quadrature (gray).  Measurement at $\phi=45^{\circ}$ at frequency $2(\omega-\omega_m)/\Gamma=-5$ (square filled on left, dashed-red line) results in larger imprecision, and at frequency $2(\omega-\omega_m)/\Gamma=5$ (square filled on right, solid red line) results in reduced imprecision. The dashed black line is the QL for $2(\omega-\omega_m)/\Gamma=5$ evaluated at quantum efficiency $\epsilon=0.35$, plus thermal contribution due to $n_{th}=1.29$.  Red and blue lines are full power-dependent expectation for corresponding data points.  Inset: measured PSD as function of frequency $2(\omega-\omega_m)/\Gamma$, at $\phi=90^{\circ}$ (blue) and $\phi=45^{\circ}$ (red) quadratures for power $p = 14$, as indicated by the dotted box in main figure.  For this choice of $\phi$ negative detunings (red dashed) yield worse sensitivity, while positive detunings (solid red) are improved. In both (a) and (b), the gray asterisks are SN measurement at $\phi=90^{\circ}$ for $\epsilon=0.35$, and the dashed gray line is for $\epsilon=1$.}
	\label{fig: meas noise}
\end{figure*}

However, if the homodyne detector is arranged to measure a finite quadrature angle $0^{\circ}<\phi< 90^{\circ}$ the cross-correlation term $S_{IF}(\phi)$ in Eq.~(\ref{eq: pos spec}) becomes nonzero and $S_{xx}$ can be smaller than the SQL (Fig.~\ref{fig: exp setup}(a,b) dashed red line)~\cite{Kimble(01)_GI_inter,Clerk(10)_QN}. Because $S_{IF}$ is real, an additional requirement is that $\tilde{\chi}_m(\omega)$ has an imaginary part, which only happens off resonance.  In particular, $\Im[\tilde{\chi}_m(\omega) S_{IF}(\phi)]$ gives rise to a Fano-like frequency dependence, in analogy to that observed in ponderomotive squeezing of light.  While such squeezed light has been observed \cite{Brooks(12)_Squeez,SafaviNaeini(13)_Squeezing,Purdy(13)_MechSqueeze}, improved measurements were far from accessible in previous experiments due to technical noise at large $\phi$. Figure~\ref{fig: exp setup}(d) shows for comparison ponderomotive squeezing of the cavity output light.  The optimal measurement phase ($\phi_{\rm{opt}}$) is rotated towards PM  compared to the optimal squeezing phase.  While rotating towards AM introduces the correlations of interest it also dilutes mechanical information found only in PM (see Appendix).  In variational readout, $\phi$ is tuned as a function of frequency to approach the QL over a broader range of frequencies~\cite{Kimble(01)_GI_inter}.  Variational readout at a fixed power is illustrated by the red line of Fig.~\ref{fig: exp setup}(a).  As seen in Fig.~\ref{fig: exp setup}(b) the power can be optimized in order to reach the QL (solid black line), which corresponds to total noise at twice the zero point motion.

Variational readout in a homodyne measurement is a technique most suited to broadband, off-resonance displacement or force measurement for $2\abs{\omega-\omega_m}/\Gamma \ge 1$.  However, even for on-resonance force measurements quantum correlations can be utilized by employing a two-tone local oscillator, using recently proposed synodyne readout~\cite{Buchmann(16)_Synodyne}.  Synodyne realizes single-quadrature measurement at a given frequency within a range of near-resonant frequencies ($2\abs{\omega-\omega_m}/\Gamma \le 1$). (See the Supplemental Material for an analysis of synodyne and its comparison to the SQL in a frequency-domain picture analogous to Fig.~\ref{fig: exp setup}.)  Variational readout and synodyne are related in that they both approach the readout problem by modifying the local oscillator, instead of for example the intracavity field.  Lastly, note asymmetric lineshapes associated with cross-correlations have been recently observed in \cite{Clark(16)_SqueezedLightRPSN} using an input quadrature squeezed probe in the microwave domain, but an improvement in the off-resonance sensitivity has not been shown to date. Squeezed light can also be used independent of $S_{IF}$ to modify $\hat{I}_{\phi}$ or $\hat{F}_{\rm{ba}}$ to modify power requirements~\cite{Schnabel(10)_LIGOsque, Kimble(01)_GI_inter}.  Injecting squeezed vacuum into the dark port of an interferometer is a technique that has been already implemented in large-scale interferometers (such as advanced LIGO), but their aim was to enable better sensitivity in a fully shot-noise limited band, without increasing optical power~\cite{LIGO(11)_LIGOsque}.

\begin{figure*}[]
	\centering
    \includegraphics[width=\textwidth,trim={0.25cm 20.5cm 4.25cm 0.5cm},clip]{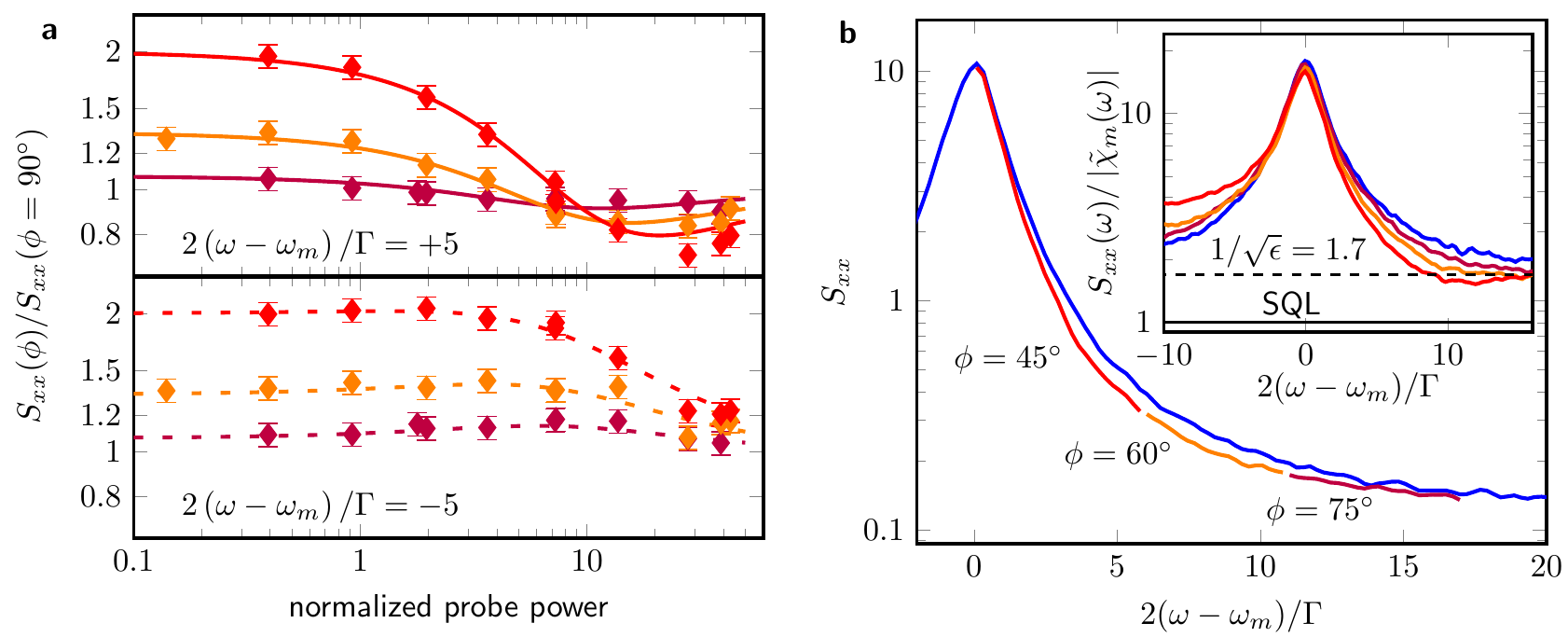} 
	\caption{Components of variational readout.  (a) Ratio of PSD in quadratures $\phi=45^{\circ}$ (red), $\phi=60^{\circ}$ (orange), and $\phi=75^{\circ}$ (purple) to that in $\phi=90^{\circ}$ as a function of normalized probe power p. The top panel is at a detuning of $2(\omega-\omega_m)/\Gamma=5$, and the bottom panel at $2(\omega-\omega_m)/\Gamma=-5$. (b) Reconstruction of variational readout for $p=14$. The measured quadratures (same colors as (a)) are placed in the portion of the spectrum for which they provide the lowest imprecision.  For comparison is the result for $\phi=90^{\circ}$ (blue).  Inset: Measured quadratures (same colors as (b)) at $p = 28$ normalized to the corresponding SQL at each frequency.  }
	\label{fig: ratio}
\end{figure*}

In our experiments, we create an optical interferometer consisting of a cryogenically-compatible Fabry-Perot cavity coupled to a high stress $\rm{Si}_3\rm{N}_4$ membrane resonator  \cite{Thompson(08)_MemInMid,Purdy(13)_MechSqueeze,Peterson(16)_BALcool} [Fig.~\ref{fig: exp setup}(c)].   We probe the motion of the $(2,2)$ membrane mode at $\omega_m/2\pi = 1.596\:\unit{MHz}$ (twice the fundamental frequency), with an exceptionally high quality factor of $Q_m=32\times10^6$. We optically damp and cool to a effective linewidth of $\Gamma/2\pi=340\:\unit{Hz}$ by injecting a tone red-detuned of the optical cavity.  We linearly monitor the displacement for the science described in this work by injecting an on-resonant probe into the cavity.  To yield a mechanical spectrum with minimal thermal noise, even a number of mechanical linewidths off resonance, we precool the cavity and membrane to $120\:\unit{mK}$, and shield the membrane mode by embedding it in a phononic crystal~\cite{Yu(14)_PnC,Tsaturyan(14)_pnc}.    At our chosen damping level, we measure via sideband asymmetry a phonon occupation of $n_{\rm{th}}=1.29\pm0.05$; this final phonon occupation is defined by the optical damping rate we choose, and is not limited by added noise~\cite{Peterson(16)_BALcool} (see Supplemental  Material). The outgoing probe light is measured using balanced homodyne detection, with a total quantum efficiency of $\epsilon=0.350\pm0.015$. The single-photon coupling was independently calibrated and found to be consistent with a value of $g/2\pi=39\:\unit{Hz}$ tightly bounded by the experimental data.

We start by measuring $S_{xx}$ on/off-resonance at $\phi=90^{\circ}$ in which the measurement noise is similar to the SQL [Fig.~\ref{fig: meas noise}(a)].  For all frequencies the relative fraction of measurement noise to the SQL at that frequency is constant and equal to $1.7$ as set by our quantum efficiency.  For the on-resonance measurements (triangles) a total measured PSD of $5.3\pm0.2$ times the on-resonance SQL added noise ($S_{\rm{sql}}$) is realized, corresponding to the smallest reported value to our knowledge
\cite{Schreppler(14)_SQL,Suh(14)_MechBAE,LaHaye(04)_appQL}.  This is due to our low phonon occupation and high quantum efficiency.  When we examine the data at $2(\omega-\omega_m)/\Gamma=10$ (circles), the total measured noise reduces to twice the off-resonant SQL value because the thermal disturbance component ($S_m$) drops faster than the SQL [Fig.~\ref{fig: exp setup}(a)].

When measuring at a finite intermediate angle ($\phi=45^{\circ}$) a distinct Fano-like lineshape due to the cross-correlations appears (inset of Fig.~\ref{fig: meas noise}(b)). We see the sensitivity is increased over a range of frequencies off resonance.  Figures~\ref{fig: meas noise}(b) and \ref{fig: ratio}(a) show how a $\phi=45^{\circ}$ measurement (red) results in an imprecision below that at $\phi=90^{\circ}$ (blue) for frequencies above resonance and near the off-resonant SQL power. On the other hand at low powers ($p \ll 1$) there is no improvement because the SN contribution increases as one adds more AM noise that doesn't contain information about the resonator, and at high powers ($p \gg 1$) the backaction contribution is dominant over the correlation and there is also no improvement.

In Fig.~\ref{fig: meas noise}(b) we see the clear measurement improvement when detecting at ($\phi=45^{\circ}$), and that it's optimal power is at a higher power than for the standard measurement configuration ($\phi=90^{\circ}$).  For a given mechanical detuning ($\rho=2(\omega-\omega_m)/\Gamma$) we find the optimal measurement quadrature to be $\cot\phi_{\rm{opt}} = \epsilon p \rho \abs{\tilde{\chi}_m(\rho)}^2$, which depends on both the quantum efficiency and measurement strength ($p$).  As derived in the Appendix, the achievable limit at the optimal power ($p_{\rm{opt}}$) depends on the quantum efficiency according to:
\begin{align}
S_{xx}(\rho,\phi_{\rm{opt}},p_{\rm{opt}}) = &
2\left(n_{\rm{th}}+\frac{1}{2}\right) \abs{\tilde{\chi}_m(\rho)}^2 + \nonumber \\
 & \sqrt{ \frac{1}{\epsilon} + \frac{1-\epsilon}{\epsilon}\rho^2 } \abs{\tilde{\chi}_m(\rho)}^2. \label{eq: Sxx QL}
\end{align}
Which is shown as the dashed black line in Fig.~\ref{fig: meas noise}(b) for $\rho=5$.  The above result at unit quantum efficiency reproduces the expected QL at all frequencies, as the amount of added noise is equal to the zero-point motion PSD contribution ($\abs{\tilde{\chi}_m(\rho)}^2$). In comparison a typical measurement is at $\phi=90^{\circ}$ (PM) and will scale with quantum efficiency as $\abs{\tilde{\chi}_m(\rho)}/\sqrt{\epsilon}$.  Note that at unit efficiency this is the difference between QL and SQL.

In variational readout, the analysis quadrature would be changed as function of frequency to realize an optimal measurement at all frequencies (reaching the QL at a single frequency). In our work, while we do not vary the quadrature in a single measurement, we are able to reconstruct variational readout spectra from four different measurements at different quadratures with the same power [Fig.~\ref{fig: ratio}(b)]. In Fig.~\ref{fig: ratio}(b)-inset we show the data normalized to the corresponding off-resonant SQL at each frequency ($S_{\rm{sql}}(\omega)$) for an even larger probe power ($p=28$), and can quantitatively analyze our improvement compared to the SQL, under the constraint of finite quantum efficiency.  At a frequency $2(\omega-\omega_m)/\Gamma=12$, we find $1.55\pm0.07$ times the SQL value for $\phi=45^{\circ}$. Taking into account our finite quantum efficiency, the minimum added noise we could hope to achieve for a $\phi=90^{\circ}$ measurement is $1/\sqrt{\epsilon}=1.7$ times the SQL value, and hence measurement at $\phi=45^{\circ}$ allows us to measure at $0.91\pm0.04$ times the finite quantum efficiency SQL.  At this frequency the thermal and zero point motion contribution of $0.3$ (in units of off-resonance $S_{\rm{sql}}$) is small, compared to the added noise due to the probe of $1.25$.

The technique we demonstrate shows the value and simplicity of utilizing imprecision-backaction correlations when carrying out strong measurement.  In this technique the degree to which the SQL can be surpassed is greatly dependent on the quantum efficiency of the probe.  In the future, pursuing higher quantum efficiency will be a natural goal of a variety of detectors, and hence extend the utility of variational readout.  High quantum efficiency combined with the ability to arbitrarily manipulate the local oscillator and corresponding correlations will offer useful opportunities to advance broadband displacement and force sensing beyond the standard quantum limit.

This work was supported by AFOSR PECASE, ONR DURIP, AFOSR-MURI, RAFAEL, the Cottrell Scholar's program, and the National Science Foundation under grant number $1125844$. We thank W. P. Bowen, T. P. Purdy, O. Wipfli, J. D. Teufel, and A. Higginbotham for useful conversation and expertise.

\appendix*
\section{Derivations}

\subsection{Optomechanical interaction}

The optomechanical interaction in the interferometer is defined by the following Hamiltonian \cite{Boerkje(10)_Observa_RPSN_Theory,Aspelmeyer(14)_RevModPhys}.
\begin{equation}
\hat{H}_0=\frac{1}{4}\hbar \omega_m \left(\hat{x}^2+\hat{p}^2 \right) + \hbar \omega_c \hat{a}^{\dag}\hat{a}+\hbar g \hat{x} \hat{a}^{\dag} \hat{a}
	\label{eq:Hamiltonian}
\end{equation}
where $\omega_m$ is the mechanical resonance frequency, $\hat{x}$ ($\hat{p}$) is the mechanical position (momentum) fluctuation operator, $\omega_c$ is the optical cavity resonance frequency, $\hat{a}$ $(\hat{a}^{\dag})$ is the optical intracavity annihilation (creation) operator,  $g$ is a single-photon optomechanical coupling constant.  In this way of writing the Hamiltonian the position and momentum operators are normalized to their zero-point fluctuations $x_{\mathrm{zp}}=\sqrt{\hbar/2 m \omega_m}$ and $p_{\rm{zp}}=\sqrt{\frac{1}{2}m\hbar\omega_m}$.  Here $m$ the mechanical resonator effective mass, and $\hbar$ the reduced Planck constant.  In our analysis the Hamiltonian is linearized by assuming a large optical coherent state.  Thus we write the optical annihilation operator as $\hat{a} = \bar{a} + \hat{u}$ with $\bar{a} = \langle \hat{a} \rangle$, and neglecting the $\hat{u}^{\dagger} \hat{u}$ term.

\vspace{-12 pt}
\subsection{Heisenberg-Langevin equations for the light operators} \label{light operators}

Here we present the analysis of the Heisenberg-Langevin equations for the probe light operator $u$, based on the Hamiltonian given above.  Similar to previous analyses we write the solution to the Heisenberg-Langevin equations of motion of our optomechanical system~\cite{Boerkje(10)_Observa_RPSN_Theory,Purdy(13)_MechSqueeze,Weinstein(14)_ObserInterpCool,Botter(12)_LinAmpModel}.  In subsequent sections we will convert the solutions to the units used in our final equations.  In the Supplemental Material we add to these equations a treatment of potential classical noise terms (and find them to be negligible for our experimental parameters).

The probe light quadratures $u_{AM}$ and $u_{PM}$ are given by,
\begin{align}
\left(
\begin{array}{c}
  \hat{u}_{AM}(\omega) \\
  \hat{u}_{PM}(\omega)
\end{array} \right)
&=
\left( \begin{array}{c}
  \hat{\mu}_{AM}(\omega) \\
  \hat{\mu}_{PM}(\omega)
\end{array} \right) \nonumber \\
&+\sqrt{\epsilon\kappa} g\bar{a} \left( \begin{array}{cc}
  \pi_-(\omega) & 0 \\
  0 & \pi_+(\omega)
\end{array} \right) \left( \begin{array}{c}
  \hat{x}(\omega) \\
  \hat{x}(\omega)
\end{array} \right)
  \label{eq: u mat}
\end{align}
here $\hat{\mu}_{AM}(\omega)$ and $\hat{\mu}_{PM}(\omega)$ are the light shot-noise Langevin operators for the amplitude (AM, $\phi=0^{\circ}$) and phase (PM, $\phi=90^{\circ}$) quadratures, $\bar{a}$ is the intracavity coherent state amplitude, $\epsilon$ is the quantum efficiency, $\hat{x}(\omega)$ is the resonator state, and $\pi_{\pm}(\omega)$ are the constructive and destructive cavity susceptibility interference functions defined as $\pi_+(\omega) = \chi^*_c(-\omega) + \chi_c(\omega)$ and $\pi_-(\omega) = i\left ( \chi^*_c(-\omega) - \chi_c(\omega) \right)$.  $\chi_c(\omega)$ is the cavity susceptibility given by $\chi_c(\omega) = \left( \kappa/2 -i(\omega+\Delta)\right)^{-1}$, with $\Delta$ the probe detuning relative to the cavity resonance, and $\kappa$ the optical cavity linewidth.

The measured light operator at a given phase $\phi$, is given by $\hat{u}_{\phi} = \hat{u}_{AM}\cos\phi - \hat{u}_{PM}\sin\phi$. We use this to calculate the symmetrized light PSD, $S_{\phi}(\omega,\phi) = \expect{\hat{u}_{\phi}(-\omega)\hat{u}_{\phi}(\omega)}$~\cite{Clerk(10)_QN,Boerkje(10)_Observa_RPSN_Theory,ClerkLesHouches}. Then the dimensionless $S_{\phi}$ is decomposed as follows,
\begin{align}
S_{\phi}(\omega,\phi) &= 1 + f_{xx}(\omega,\phi) \expect{\hat{x}\hat{x}}(\omega) + S_{\mu \hat{x}}(\omega,\phi) \label{Eqs: Sp}
\end{align}
where $1$ is the probe shot-noise, $f_{xx}(\omega,\phi)$ is the transfer function from displacement to light, $\expect{\hat{x}\hat{x}}(\omega)$ is the symmetrized resonator displacement distribution PSD, and $S_{\mu \hat{x}}(\omega,\phi)$ is the cross-correlation PSD between the light shot-noise and the resonator state as inferred by the probe.

To explicitly write the above functions we define the following cavity and light parameters: $n_{\rm{th}}$ is the thermal phonon occupation, $\Gamma$ the effective mechanical linewidth, $\omega_m$ the resonator frequency, and the mechanical susceptibility is given by $\chi_m(\omega) = \left( \Gamma/2 - i\left(\omega - \omega_m\right) \right)^{-1}$. We find:\begin{align}
f_{xx}(\omega,\phi) &= \epsilon \kappa (g\bar{a})^2 \left\{ \abs{\chi_{c}(-\omega)}^2 + \abs{\chi_c(\omega)}^2 \right\} \nonumber \\
&  - 2 \epsilon \kappa (g\bar{a})^2 \Re\left[\chi_c(-\omega)\chi_c(\omega)e^{-2i\phi} \right]  \label{Eqs: fxx full} \\
S_{\mu \hat{x}}(\omega,\phi) &= \epsilon \kappa \left(g\bar{a}\right)^2 \left( \abs{\chi_{c}(-\omega)}^2 - \abs{\chi_{c}(\omega)}^2 \right) \Im \left[i\chi_m(\omega) \right] \nonumber \\
&-2\epsilon \kappa (g\bar{a})^2 \Im\left[ \chi_c(-\omega)\chi_c(\omega)e^{-2i\phi} \right]\Re\left[ i\chi_{m}(\omega) \right]
\end{align}
\begin{align}
\expect{\hat{x}\hat{x}}(\omega) &= \Gamma(n_{\rm{th}}+1/2) \abs{\chi_m(\omega)}^2  \nonumber \\
&+ (g\bar{a})^2\abs{\chi_m(\omega)}^2 \frac{\kappa}{2}\left(\abs{\chi_c(-\omega)}^2 + \abs{\chi_c(\omega)}^2 \right) \nonumber \\
&+ \left(\frac{F}{4p_{\rm{zp}}}\right)^2 \abs{\chi_m(\omega)}^2 \delta(\omega-\omega_f)  \label{Eqs: xx full}
\end{align}
In Eq.~(\ref{Eqs: xx full}) we have included the response to external force ($F$), applied on the resonator at a frequency $\omega_f$.  Note that the $\Gamma$ and $\omega_m$ are effective mechanical parameters due to well-known optical damping and spring effects dominantly from a red-detuned damping tone but with a very small contribution from the small detuning of the probe~\cite{Aspelmeyer(14)_RevModPhys}.

\vspace{-12 pt}
\subsection{Derivation of the standard quantum limit value} \label{sec: sql}

Here we derive the SQL PSD ($S_{\rm{sql}}$) using the result above by minimizing the combined shot-noise and quantum backaction terms~\cite{Braginsky(03)_NonRelaventSQL}.
\begin{align}
S_{\rm{sql}}(\omega) = x_{\rm{zp}}^2
\min_{\bar{a},\phi,\Delta} \{ \frac{1}{f_{xx}(\omega,\phi)} + \nonumber \\
(g\bar{a})^2\abs{\chi_m(\omega)}^2 \frac{\kappa}{2}\left(\abs{\chi_{c}(-\omega)}^2 + \abs{\chi_{c}(\omega)}^2 \right) \} \label{eq: ssql min func}
\end{align}
with $f_{xx}(\omega,\phi)$ the transfer function from displacement to light (with units of hertz) of Eq.~(\ref{Eqs: fxx full}).
The $S_{\rm{sql}}$ is composed only of the added noise by the measurement probe (shot-noise and quantum backaction), and ignores the mechanical state (thermal and zero-point motion). This is motivated from the point of view that the mechanical resonator state is the signal to measure. We note that in Ref.~\cite{Schreppler(14)_SQL}, the zero-point motion is included in the total noise.

The minimum value of the SQL PSD is found for detection angle of $\phi=90^{\circ}$ and for probing on cavity resonance ($\Delta=0$). Additionally we evaluate $f_{xx}$ for unity detection efficiency ($\epsilon=1$). With this we find the number of photons required for SQL detection to be,
\begin{align}
\bar{a}^2_{\rm{sql}}(\omega) =& \frac{1}{2\kappa g^2 \abs{\chi_c(\omega)}^2 \abs{\chi_m(\omega)} }  = \nonumber \\
& \frac{\Gamma}{4\kappa g^2 \abs{\chi_c(\omega)}^2 } \sqrt{1 + \rho^2 } \label{eq: a-sql}
\end{align}
with $\rho = 2 \left(\frac{\omega-\omega_m}{\Gamma}\right)$ a dimensionless mechanical detuning.  We combine this result with Eq.~(\ref{eq: ssql min func}) to find,
\begin{align}
S_{\rm{sql}}(\omega) = x_{\rm{zp}}^2\abs{\chi_m(\omega)} = x_{\rm{zp}}^2\frac{2}{\Gamma} \frac{1}{\sqrt{1 + \rho^2}}. \label{eq: Ssql}
\end{align}
When evaluated on-resonance the SQL value is $S_{\rm{sql}}(\omega_m) = \frac{\hbar}{m\omega_m\Gamma}$.  We utilize a probe power ($p$) normalized to the on-resonant SQL power, i.e. $p=\frac{\bar{a}^2}{\bar{a}_{\rm{sql}}^2(\omega_m)}$.  Note $p$ is related to the commonly-used optomechanical cooperativity ($C$) by $C=p/4$~\cite{Schreppler(14)_SQL,Aspelmeyer(14)_RevModPhys} .

\vspace{-12 pt}
\subsection{Converting measurement noise results to dimensionless displacement PSD} \label{sec: heis}

In this section we write the result of the Heisenberg-Langevin equations for the light operators inferred as displacement of the resonator and use it to evaluate the displacement PSD.  While the light equations and associated PSD have been derived many times in the literature~\cite{Boerkje(10)_Observa_RPSN_Theory,Purdy(13)_MechSqueeze,Weinstein(14)_ObserInterpCool,Botter(12)_LinAmpModel}, our approach is to use quantum noise notation, as described in for example \cite{Clerk(10)_QN}, which allows us to see more transparently how the QL can be reach for a particular optical probe configuration.

The probe light operator converted to inferred displacement $\hat{X}$ is given by multiplying the probe light operator $\hat{u}$ by $x_{\rm{zp}}/\sqrt{f_{xx}(\phi=90^{\circ})}$($=x_{\rm{zp}}/\sqrt{\epsilon \Gamma p}$), which gives dimensions of $\unit{m/\sqrt{Hz}}$.  We normalize the displacement by the on-resonance SQL amplitude ($\sqrt{S_{\rm{sql}}(\omega_m)}$). We also use a dimensionless mechanical susceptibility defined by $\tilde{\chi}_m(\rho) = \chi_m(\omega)/\abs{\chi_m(\omega_m)}= (1 - i \rho)^{-1}$ and a dimensionless cavity susceptibility $\tilde{\chi}_c(\omega) = \chi_c(\omega)/\abs{\chi_c(\omega_m)}$. With this we write the AM and PM quadrature operators for the inferred displacement as,
\begin{align}
\hat{X}_{AM}(\omega) &= \frac{1}{\sqrt{2\epsilon p}\tilde{\chi}_c(\omega)} \hat{\mu}_{AM}(\omega) \\
\hat{X}_{PM}(\omega) &= \frac{1}{\sqrt{2\epsilon p}\tilde{\chi}_c(\omega)} \hat{\mu}_{PM}(\omega) + \sqrt{\frac{\Gamma}{2}} \hat{x}(\omega) \\
\sqrt{\frac{\Gamma}{2}} \hat{x}(\omega)  &= \hat{x}_{m}(\omega) - 2\omega_m\chi_m^*(-\omega)\tilde{\chi}_m(\omega) \hat{F}_{\rm{ba}}(\omega)
\end{align}
with $\hat{x}_{m}$ the resonator state (which includes zero-point motion, thermal and external force), and $\hat{F}_{\rm{ba}}$ the dimensionless backaction force applied onto the resonator. The backaction operator is proportional to $\sqrt{p}$ and is a function of the shot-noise AM quadrature operator~\cite{Boerkje(10)_Observa_RPSN_Theory,Purdy(13)_MechSqueeze}.

Because the resonator ($\hat{x}$) information is only in the PM quadrature, the displacement phase dependence is given by,
\begin{align}
\hat{X}_{\phi}(\omega) &= -\cot\phi \hat{X}_{AM}(\omega) + \hat{X}_{PM}(\omega) \nonumber \\
                       &= \hat{x}_{m}(\omega) + \hat{I}_{\phi} - 2\omega_m\tilde{\chi}_m^*(-\omega)\tilde{\chi}_m(\omega) \hat{F}_{\rm{ba}}(\omega) \nonumber \\
                       &\cong \hat{x}_{m}(\omega) + \hat{I}_{\phi} - i\tilde{\chi}_m(\omega) \hat{F}_{\rm{ba}}(\omega) \label{eq: x phi}
\end{align}
Here we have defined the shot-noise displacement imprecision operator to be $\hat{I}_{\phi} = \frac{1}{\sqrt{2\epsilon p}\tilde{\chi}_c(\omega)} \left(-\cot\phi \hat{\mu}_{AM}(\omega) + \hat{\mu}_{PM}(\omega)\right)$, from which it follows that there is a weak frequency dependence proportional to $\abs{1/\tilde{\chi}_c(\omega)}$.  In the third line we take the limit of a high-Q resonator ($\Gamma\ll \omega_m$).  Note in AM $\cot\phi = \cot 0 = \infty$ corresponds to the case in which all the information about the resonator is in the $\phi=90^{\circ}$ (PM) quadrature and thus the displacement measurement diverges. It follows that the displacement PSD is given by,
\begin{align}
S_{xx}(\omega) &= \expect{\hat{X}_{\phi}(-\omega)\hat{X}_{\phi}(\omega)} \nonumber \\
&\cong S_{m}(\omega) + S_{II}(\omega) + \abs{\tilde{\chi}_m(\rho)}^2 S_{FF}(\omega) + \nonumber \\
&2 \Im \left[ \tilde{\chi}_m(\rho) S_{IF} \right] \label{eq: sxx heis}
\end{align}
with functions given by,
\begin{align}
S_{m}(\omega) &= 2\left(n_{\rm{th}}+\frac{1}{2}\right) \abs{\tilde{\chi}_{m}(\rho)}^2 \\
S_{II}(\omega) &= \frac{1 + \cot^2\phi}{2 \epsilon p \abs{\tilde{\chi}_c(\omega)}^2 } \\
S_{FF}(\omega) &= \frac{1}{2} p \abs{\tilde{\chi}_c(\omega)}^2 \\
S_{IF}(\omega) &=-\frac{1}{2} \cot\phi.
\end{align}

Here $S_m$ is the displacement PSD of the resonator including thermal, and zero-point motion, in which we set the external force to zero; $S_{II}$ is the displacement imprecision PSD due to the probe SN and is inversely proportional to the power; $S_{FF}$ is the displacement change due to the backaction force; $S_{IF}$ is the cross-correlation term measured in homodyne detection and is real, but in general it can be complex \cite{Clerk(10)_QN}. For example, in synodyne detection the equivalent $S_{IF}$ is complex and $S_{IF} = -S_{FI}^*$ \cite{Buchmann(16)_Synodyne}.

When examining the contribution of the backaction in Eq.~(\ref{eq: x phi}), there is a $90^{\circ}$ degrees phase delay with respect to the shot-noise term ($\hat{I}_{\phi}$). This means there is a time lag between the backaction force applied on the resonator and the measurement time. For this reason, there must be a frequency dependence in the cross-correlation term. Only the imaginary part of the mechanical susceptibility $\Im \tilde{\chi}_m(\rho) = \rho\abs{\tilde{\chi}_m(\rho)}^2$ contributes to the PSD.

It will also be useful to explicitly write out the PSD with the assumption $\abs{\tilde{\chi}_c(\omega)}^2 = 1$, which is a very good approximation for our experimental parameters.
\begin{align}
S_{xx}(\omega,p,\phi)
&= 2\left(n_{\rm{th}}+\frac{1}{2}\right) \abs{\tilde{\chi}_{m}(\omega)}^2 + \frac{1 + \cot^2\phi}{2 \epsilon p} + \nonumber \\
&\frac{1}{2} p \abs{\tilde{\chi}_{m}(\omega)}^2
- \cot\phi \rho \abs{\tilde{\chi}_m(\omega)}^2 \label{eq: Sx reso}
\end{align}

\vspace{-24 pt}
\subsection{Comparison to uncertainty relations and parameters for reaching the QL}

Through Eq.~(\ref{eq: sxx heis})-(\ref{eq: Sx reso}) we can compare the PSD to the Heisenberg uncertainty relation \cite{Robertson(34)_statUncePrinc} for the state of probe light and to the SQL. The Heisenberg uncertainty relation is given by,
\begin{align}
\Delta \hat{I}_{\phi}^2\Delta \hat{F}_{\rm{ba}}^2 &\ge \frac{1}{4} \abs{ \expect{\left[\hat{I}_{\phi},\hat{F}_{\rm{ba}} \right]} }^2 + \frac{1}{4} \abs{ \expect{\left\{\hat{I}_{\phi},\hat{F}_{\rm{ba}} \right\}} }^2 \nonumber \\
S_{II} S_{FF} &\ge \frac{1}{4}  + S_{IF}^2
\end{align}
For $\epsilon=1$ and $\phi = 90^{\circ}$, $S_{II} S_{FF} = \frac{1}{4}$, which is the case for an SQL measurement configuration, and is power independent. But when measuring at some intermediate angle ($0^{\circ} < \phi < 90^{\circ}$), the measurement imprecision can be below the SQL value of $1/4$.

We can also determine the correct choice of $p$ and $\phi$ to reach the QL, when the additional contribution of the cross-correlation term ($S_{IF}$) is taken into account. When we optimize the PSD (Eq.~(\ref{eq: Sx reso})) to find the optimal measurement phase we find $\cot\phi_{\rm{opt}} = \epsilon p \rho \abs{\tilde{\chi}_m(\rho)}^2$. Placing this back into Eq.~(\ref{eq: Sx reso}) we find the succinct consequence of variational readout,
\begin{align}
S_{xx}(\rho,\phi_{\rm{opt}},p) =
2\left(n_{\rm{th}}+\frac{1}{2}\right) \abs{\tilde{\chi}_m(\rho)}^2 + \nonumber \\
\frac{1}{2 \epsilon p } + \frac{1}{2} p \left( 1 + (1-\epsilon)\rho^2 \right) \abs{\tilde{\chi}_m(\rho)}^4 \label{eq: Sx-sql-opt}
\end{align}
for a given QE.

At the optimal power $p_{\rm{opt}}$ and optimal angle one finds precisely the QL:
\begin{align}
S_{xx}(\rho,\phi_{\rm{opt}},p_{\rm{opt}}) =
2\left(n_{\rm{th}}+\frac{1}{2}\right) \abs{\tilde{\chi}_m(\rho)}^2 + \nonumber \\
\sqrt{ \frac{1}{\epsilon} + \frac{1-\epsilon}{\epsilon}\rho^2 } \abs{\tilde{\chi}_m(\rho)}^2 \label{eq: Sxx QL} \\
p_{\rm{opt}} = \frac{1}{\sqrt{ \epsilon \left( 1 + (1-\epsilon)\rho^2 \right) } \abs{\tilde{\chi}_m(\rho)}^2 }
\end{align}
We recall that the QL for an ideal detector reaches one zero-point motion contribution for each measured mechanical quadrature [black lines in Fig.~\ref{fig: exp setup}(a)].

\vspace{-12 pt}
\subsection{Comparison to ponderomotive squeezing} \label{sec: comp sque}

The above analysis gives us the necessary tools to directly compare the consequence of variational readout on displacement sensitivity to the creation of ponderomotive squeezing.  Namely, we can compare the $S_{\phi}$ derived in Sec.~\ref{light operators} to $S_{xx}$ to find the light PSD has a different $\phi$ dependence than the displacement PSD~\cite{Brooks(12)_Squeez,SafaviNaeini(13)_Squeezing,Purdy(13)_MechSqueeze}. We can write this as,
\begin{align}
S_{\phi}(\rho,\phi,p) = 2\epsilon p \sin^2\phi S_{xx}(\rho,\phi,p) .
\end{align}
The different phase dependences is illustrated in the main text in Fig.~\ref{fig: exp setup}(d).  For light squeezing the information about the resonator does not matter. Conversely, for displacement measurement, while the larger cross-correlation term reduces the backaction contribution by rotating towards the AM quadrature, it also dilutes mechanical information found in the PM quadrature. For this reason the optimal ponderomotive squeezing angle is closer to the AM quadrature than for the displacement measurement.


%

\widetext
\newpage

\begin{center}
\textbf{\large Supplemental Materials: \\ Improving broadband displacement detection with quantum correlations
}
\end{center}
\setcounter{equation}{0}
\setcounter{figure}{0}
\setcounter{table}{0}
\setcounter{page}{1}
\setcounter{section}{0}
\makeatletter
\renewcommand{\theequation}{S\arabic{equation}}
\renewcommand{\thesection}{S\arabic{section}}
\renewcommand{\thefigure}{S\arabic{figure}}

\section{Overview of Experimental Protocol and Noise Considerations} \label{sec: overview}

Our optical cavity has a linewidth of $\kappa/2\pi=2.50\pm 0.16$ $\unit{MHz}$, and is frequency-stabilized (locked) with respect to a weak on-cavity-resonance beam. A second damping beam, with the same polarization, is red detuned of the cavity resonance by 2 $\unit{MHz}$ and continuously cools the membrane. Lastly, a coherent on-cavity-resonance beam, with orthogonal polarization to both the lock and damping beams, is used as the measurement probe beam. This choice minimizes the cross talk between the probe beam and the other beams.  We do not observe a relevant classical laser noise contribution on the probe (Sec.~\ref{sec: class noise exp}) \cite{Peterson(16)_BALcool}, nor technical noise due to mechanical bath at this damping choice, nor laser heating at a $120\:\unit{mK}$ dilution refrigerator temperature (Sec.~\ref{sec: nth calib}).  We measure the probe beam via balanced detection using an external local oscillator. We set the detection quadrature phase by locking the relative phase between the local oscillator and the probe beam. The error signal uses the DC signal from a balanced homodyne receiver to realize measurement phases $40^{\circ}\le\phi\le 140^{\circ}$, which are calibrated based upon full excursion of this error signal.
From the time trace of the probe output we compute the symmetrized power spectral density (PSD), and average over $1000-7500$ traces for each measurement. Then we normalize to a trace taken without the probe beam to achieve a shot-noise normalized frequency spectrum $\expect{\hat{u}_{\phi}(-\omega)\hat{u}_{\phi}(\omega)}$.
To calibrate the PSD we require three parameters:  The SQL power determined from the coupling coefficient $g$ (Sec.~\ref{sec: g calib}), phonon occupation $n_{\rm{th}}$ (Sec.~\ref{sec: nth calib}), and quantum efficiency $\epsilon$ (Sec.~\ref{sec: eps calib}).

\section{Optomechanical coupling coefficient calibration} \label{sec: g calib}

We calibrate the single photon optomechanical coupling coefficient $g$ by determining the phonon occupation via sideband asymmetry and combining this with independent knowledge of the number of cooling photons in the cavity $N_{\rm{damp}}$.  To extract the phonon occupation via sideband asymmetry, we use a red detuned damping beam, and measure the red and blue sideband amplitudes as a function of optical power as described in~\cite{Peterson(16)_BALcool}. To determine $N_{\rm{damp}}$ we measure the power that leaves the cavity and convert it to cavity photon number, which requires knowledge only of the cavity linewidth and the detuning of the damping beam.  $g$ is then determined via
\begin{align}
g^2 N_{\rm{damp}} = \frac{A_{b}}{\epsilon \kappa \abs{\chi_c(\omega_m)}} \frac{\Gamma}{4n_{\rm{th}}}
\end{align}
Here $A_{b}$ is the blue sideband amplitude, $n_{\rm{th}}$ is the phononic occupation determined from the sideband asymmetry, $\epsilon$ is the quantum efficiency, $\chi_c(\omega_m)$ is the on-mechanical-resonance cavity susceptibility, and $\kappa$ is the cavity linewidth.

\begin{figure}[b]
	\centering
    \includegraphics[trim={1.5cm 20.5cm 2.25cm 0.25cm},clip]{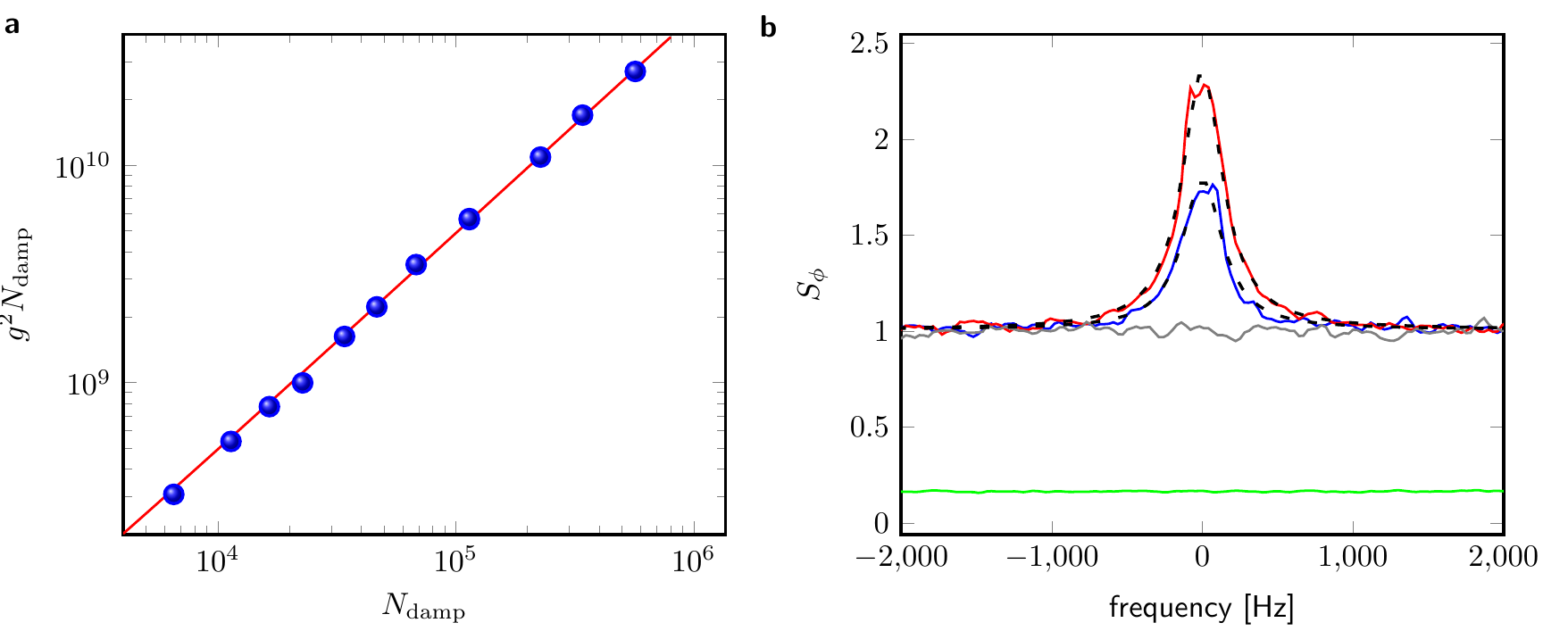}  
	\caption{(a) Measurements of the coupling coefficient as a function of the number of photons inside the cavity. The red line is a linear fit to the data. (b) The red and blue sideband shot-noise normalized PSD in balanced heterodyne detection average over all phases of an on-resonance probe, used to find the phonon occupation. The gray curve is the shot-noise PSD and the green curve is the electronic noise PSD. The dashed black lines are fits to the data, and the fit parameters are given in the text.}
	\label{fig: calib}
\end{figure}

In Fig.~\ref{fig: calib}(a) we plot the evaluated $g^2 N_{\rm{damp}}$ as function of the number of damping photons in the cavity, to determine $g/2\pi =  35.0\pm 5.5\:\unit{Hz}$ from the slope.  The data in Figs.~2 and~3 of the main text constrain the coupling coefficient to $g/2\pi = 39\:\unit{Hz}$.  This absolute value of $g$ is used to define the SQL power $p$ that is important in the main text.

In these measurements we also find the backaction cooling limit to be $n_{\rm{ba}} = 0.16 \pm 0.02$~\cite{Peterson(16)_BALcool}.  This value is within the expected backaction limit of $0.15\pm0.01$ phonons.

\section{Phonon occupation calibration} \label{sec: nth calib}

While a red-detuned probe allows us to understand general cavity parameters, it is also important to understand the specific phonon occupation  $n_{\rm{th}}$ for the on-resonance-probe measurements.  For this we damp the mechanical resonator to a linewidth close to the value we used in the data of Figs.~2 and~3 of the main text, and also apply a weak on-cavity-resonance probe beam detected via balanced heterodyne averaging over all phases, to measure the red ($A_r$) and blue ($A_b$) sideband amplitudes.  The result is given in Fig.~\ref{fig: calib}(b).  We fit the measurement to a Lorentzian to find a mechanical linewidth $\Gamma = 325\pm0.01\:\unit{Hz}$, blue amplitude $A_b=0.78\pm 0.01$, and red amplitude $A_r=1.35\pm 0.01$.  From this we extract the phonon occupation ($n_{\rm{th}} = (A_r/A_b-1)^{-1}$) corresponding to $n_{\rm{th}} = 1.34\pm 0.04$.

In the data of Figs.~2 and~3 in the main text the experimental damped linewidth is slightly different, and equal to $340\:\unit{Hz}$.  To account for this we use the relation $n_{\rm{th}} = \frac{n_0\Gamma_0}{\Gamma} + n_{\rm{ba}}$ and extract the actual phonon occupation in Figs.~2 and~3.  $n_0\Gamma_0$ is a constant equal to the initial photon occupation times the undamped mechanical linewidth, and $n_{\rm{ba}}$ is the cooling backaction limit described in Sec.~\ref{sec: g calib}.  With this, we determined our thermal phonon occupation to be $1.29\pm0.05$.  This value is consistent with the the membrane having thermalized to the $T=120$ mK measured using conventional thermometry of the dilution refrigerator base plate.

\section{Quantum efficiency calibration} \label{sec: eps calib}

To calibrate the quantum efficiency ($\epsilon$), we measure optical ponderomotive squeezing \cite{Purdy(13)_MechSqueezeSupp} using a detuned probe measured on a single photodetector. By using a high-probe power deep in the radiation pressure dominated region ($\sim 80$ time the SQL power), the maximum measured squeezing dip is directly proportional to the quantum efficiency and strongly sensitive to it. With this measurement we achieve a total quantum efficiency of $\epsilon_{\rm{sq}}=26\pm1\:\%$, limited by the electronic noise of the photodetector. We decompose the quantum efficiency to $\epsilon_{\rm{sq}}=\epsilon_{\rm{e.n.}} \epsilon_{\rm{meas}}$, with the electronic noise component equal to $\epsilon_{\rm{e.n.}} = 58.5\pm0.1\:\%$. From this we find $\epsilon_{\rm{meas}} = 44\pm1\:\%$.

Next we find the quantum efficiency for the experiments in Figs.~2 and~3 of the main text, in which we used balanced homodyne detection, by comparing to the measured quantum efficiency in the squeezing experiment. As the two experiments partially share the same path and same type of photodetector, we write the quantum efficiency as, $\epsilon = \epsilon_{\rm{meas}} \epsilon_{\rm{opt}} \epsilon_{\rm{vis}}^2$. Where $\epsilon_{\rm{opt}} = 95.0\pm0.5\%$ is the additional optical losses due to different optical paths, and $\epsilon_{\rm{vis}}=92\pm1\%$ is the visibility (or mode matching) between the local oscillator and the probe beam. Accounting for these elements we find $\epsilon = 35.0\pm1.5\:\%$.

\section{Full Calculation of Light Power Spectral Density with Classical Noise} \label{sec: light psd}

\subsection{Theoretical analysis with classical noise} \label{sec: class noise theory}

In the Appendix we analyzed the Heisenberg-Langevin equations for the probe light operator.  This analysis left out potential classical noise contributions. Here we repeat this analysis and add in the classical noise fluctuations contribution.  While classical noise in our experiment is ultimately negligible, making sure measurement contributions are propagated correctly is required to come to this conclusion (see Sec.~\ref{sec: class noise exp}).  The light probe quadratures are given by,
\begin{align}
\left(
\begin{array}{c}
  \hat{u}_{AM}(\omega) \\
  \hat{u}_{PM}(\omega)
\end{array} \right)
 =
\left( \begin{array}{c}
  \hat{\mu}_{AM}(\omega) \\
  \hat{\mu}_{PM}(\omega)
\end{array} \right)
&-\sqrt{\frac{\epsilon}{2}}\kappa \left( \begin{array}{cc}
 -\pi_+(\omega) & \pi_-(\omega) \\
  \pi_-(\omega) & \pi_+(\omega)
\end{array} \right)\left( \begin{array}{c}
  \delta \tilde{y}_{AM}(\omega) \\
  \delta \tilde{y}_{PM}(\omega)
\end{array} \right) 
+\sqrt{\epsilon\kappa} g\bar{a} \left( \begin{array}{cc}
  \pi_-(\omega) & 0 \\
  0 & \pi_+(\omega)
\end{array} \right) \left( \begin{array}{c}
  \hat{x}(\omega) \\
  \hat{x}(\omega)
\end{array} \right)
  \label{eq: u mat}
\end{align}
here $\hat{\mu}_{AM}(\omega)$ and $\hat{\mu}_{PM}(\omega)$ are the light shot-noise Langevin operators for the amplitude (AM, $\phi=0^{\circ}$) and phase (PM, $\phi=90^{\circ}$) quadratures, $\bar{a}$ is the intracavity coherent state amplitude, $\hat{x}$ is the resonator state, $\delta \tilde{y}_{AM}$ ($\delta \tilde{y}_{PM}$) is the classical noise amplitude (phase) noise normalized as fraction of the shot-noise contribution in the cavity, and $\pi_{\pm}(\omega)$ are the constructive and destructive cavity susceptibility interference functions defined as $\pi_+(\omega) = \chi^*_c(-\omega) + \chi_c(\omega)$ and $\pi_-(\omega) = i\left ( \chi^*_c(-\omega) - \chi_c(\omega) \right)$; with $\chi_c(\omega)$ the cavity susceptibility. The cavity susceptibility is given by $\chi_c(\omega) = \left( \kappa/2 -i(\omega+\Delta)\right)^{-1}$, with $\Delta$ the probe detuning relative to the cavity resonance.

The measured shot-noise normalized light operator at a given phase $\phi$, is given by $\hat{u}_{\phi} = \hat{u}_{AM}\cos\phi - \hat{u}_{PM}\sin\phi$. We use this to calculate the symmetrized light PSD, $S_{\phi}(\omega,\phi) = \expect{\hat{u}_{\phi}(-\omega)\hat{u}_{\phi}(\omega)}$~\cite{Clerk(10)_QNSupp,Boerkje(10)_Observa_RPSN_TheorySupp}. Then the dimensionless $S_{\phi}$ is decomposed as follows,
\begin{align}
S_{\phi}(\omega,\phi) &= 1 + S_{\rm{LN}}(\omega,\phi) + f_{xx}(\omega,\phi) \expect{\hat{x}\hat{x}}(\omega) + S_{\mu \hat{x}}(\omega,\phi) + S_{\delta y \hat{x}}(\omega,\phi) \label{Eqs: Sp}
\end{align}
where $1$ is the probe shot-noise, $S_{\rm{LN}}(\omega,\phi)$ is the classical noise contribution, $f_{xx}(\omega,\phi)$ is the transfer function from light to displacement, $\expect{\hat{x}\hat{x}}(\omega)$ is the symmetrized resonator displacement distribution PSD, $S_{\mu \hat{x}}(\omega,\phi)$ is the cross-correlation PSD between the light shot-noise operator and the resonator state operator, and $S_{\delta y \hat{x}}(\omega,\phi)$ is the cross-correlation PSD between the light classical noise and the resonator state operator.

To explicitly write the above functions we define the following cavity and light parameters: $n_{\rm{th}}$ is the thermal phonon occupation, $\Gamma$ the mechanical linewidth, $\omega_m$ the resonator frequency, and the mechanical susceptibility is given by $\chi_m(\omega) = \left( \Gamma/2 - i\left(\omega - \omega_m\right) \right)^{-1}$. The classical noise correlations are given by $\tilde{C}_{AA}=\expect{\delta \tilde{y}_{AM}\delta \tilde{y}_{AM}}$, $\tilde{C}_{PP}=\expect{\delta \tilde{y}_{PM}\delta \tilde{y}_{PM}}$, and $\tilde{C}_{AP} = \sqrt{\tilde{C}_{AA}\tilde{C}_{PP}}$. Here the normalization for the classical noise is chosen such that it's a fraction of the in-cavity probe light shot-noise, e.g. for positive frequencies the total amplitude noise consisting of shot noise plus classical AM noise is $\frac{\kappa}{2} \abs{\chi_c(\omega)}^2 \left( 1 + \tilde{C}_{AA} \right)$. This is a convenient choice as we typically refer to the in-cavity photon number.
With this we write the light and displacement symmetrized PSD functions as:
\begin{align}
S_{\rm{LN}}(\omega,\phi) &= 2\epsilon \left(\tilde{C}_{AA} + \tilde{C}_{PP} \right) \left(\frac{\kappa}{2}\right)^2 \left(\abs{\chi_{c}(-\omega)}^2 + \abs{\chi_c(\omega)}^2 \right) \nonumber \\
&+4\epsilon\left(\tilde{C}_{AA} - \tilde{C}_{PP}\right) \left(\frac{\kappa}{2}\right)^2\Re\left[ \chi_c(-\omega)\chi_c(\omega) e^{-2i\phi} \right]  
-8\epsilon \tilde{C}_{AP} \left(\frac{\kappa}{2}\right)^2\Im\left[ \chi_c(-\omega)\chi_c(\omega) e^{-2i\phi} \right]  \label{eq: s-ln} \\
f_{xx}(\omega,\phi) &= \epsilon \kappa (g\bar{a})^2 \left\{ \abs{\chi_{c}(-\omega)}^2 + \abs{\chi_c(\omega)}^2 - 2 \Re\left[\chi_c(-\omega)\chi_c(\omega)e^{-2i\phi} \right] \right\} \label{eq: fxx full} \\
S_{\mu \hat{x}}(\omega,\phi) &= \epsilon \kappa \left(g\bar{a}\right)^2 \left( \abs{\chi_{c}(-\omega)}^2 - \abs{\chi_{c}(\omega)}^2 \right) \Im \left[i\chi_{\rm{eff}}(\omega) \right] \nonumber \\
&-2\epsilon \kappa (g\bar{a})^2 \Im\left[ \chi_c(-\omega)\chi_c(\omega)e^{-2i\phi} \right]\Re\left[ i\chi_{\rm{eff}}(\omega) \right] \\
S_{\delta y \hat{x}}(\omega,\phi) &= 
4\epsilon  (g\bar{a})^2\left(\frac{\kappa}{2}\right)^2 \left( \abs{\chi_{c}(-\omega)}^2 - \abs{\chi_{c}(\omega)}^2 \right) \left\{\tilde{C}_{AA} \Im\left[i\chi_{\rm{eff}}(\omega) \pi_+(\omega)\right] - \tilde{C}_{AP} \Im \left[i\chi_{\rm{eff}}(\omega) \pi_-(\omega)\right] \right\} \nonumber \\
&+4\epsilon (g\bar{a})^2 \left(\frac{\kappa}{2}\right)^2\left( \abs{\chi_{c}(-\omega)}^2 + \abs{\chi_{c}(\omega)}^2 \right) \left\{ \tilde{C}_{AP} \Re\left[i\chi_{\rm{eff}}(\omega)\pi_+(\omega) \right] - \tilde{C}_{PP} \Re \left[i\chi_{\rm{eff}}(\omega) \pi_-(\omega)\right] \right\} \nonumber \\
&-8\epsilon(g\bar{a})^2 \left(\frac{\kappa}{2}\right)^2\Im\left[ \chi_c(-\omega)\chi_c(\omega)e^{-2i\phi} \right] \left\{ \tilde{C}_{AA} \Re\left[i\chi_{\rm{eff}}(\omega)\pi_+(\omega)\right] - \tilde{C}_{AP} \Re\left[i\chi_{\rm{eff}}(\omega)\pi_-(\omega)\right] \right\} \nonumber \\
&-8\epsilon (g\bar{a})^2 \left(\frac{\kappa}{2}\right)^2\Re\left[ \chi_c(-\omega)\chi_c(\omega)e^{-2i\phi} \right] \left\{ \tilde{C}_{AP} \Re\left[i\chi_{\rm{eff}}(\omega)\pi_+(\omega)\right] - \tilde{C}_{PP} \Re\left[i\chi_{\rm{eff}}(\omega)\pi_-(\omega) \right] \right\}
\end{align}
\begin{align}
\expect{\hat{x}\hat{x}}(\omega) &= \Gamma(n_{\rm{th}}+1/2) \abs{\chi_{\rm{eff}}(\omega)}^2  \nonumber \\
&+ (g\bar{a})^2\abs{\chi_{\rm{eff}}(\omega)}^2 \frac{\kappa}{2}\left(\abs{\chi_c(-\omega)}^2 + \abs{\chi_c(\omega)}^2 \right) \nonumber \\
&+ (g\bar{a})^2\abs{\chi_{\rm{eff}}(\omega)}^2 \frac{\kappa}{2}\left( \abs{\pi_+}^2 \tilde{C}_{AA} + \abs{\pi_-}^2 \tilde{C}_{PP} - 4\Im\left[\chi_c(-\omega)\chi_c(\omega) \right]\tilde{C}_{AP} \right) \nonumber \\
&+ \left(\frac{F}{4p_{\rm{zp}}}\right)^2 \abs{\chi_{\rm{eff}}(\omega)}^2 \delta(\omega-\omega_f)  \label{Eqs: xx full}
\end{align}
where we have used the same effective mechanical parameters as described in the Appendix.

\subsection{The effect of classical noise in our experiments} \label{sec: class noise exp}

In this subsection we examine the effect of classical noise on the data in Figs.~2 and~3 of the main text. Specifically we are interested in quantifying the squashing effect that can rise due to the correlation between the classical noise and the mechanical resonator, i.e. $S_{\delta y \hat{x}}(\omega,\phi)$.  Other classical noise effects would only increase the measurement in Figs.~2 and~3.

We write the dimensionless displacement due to classical noise cross-correlation,
\begin{align}
\frac{S_{\delta y \hat{x}}(\omega,\phi)}{S_{\rm{sql}}(\omega_m)f_{xx}} = -\left( \cot\phi \tilde{C}_{AA} + \tilde{C}_{AP} \right) \Im \left[ \kappa \chi_c(\omega) \tilde{\chi}_m(\omega)\right]
\end{align}
Here we have used an on-resonance probe. The $\tilde{C}_{PP}$ falls out because $\abs{\chi_{c}(-\omega)}^2 - \abs{\chi_{c}(\omega)}^2 < 3\times10^{-3}$ for detuning $\Delta/2\pi<5\:\unit{kHz}$. Therefore the main contribution for the noise component comes from $\tilde{C}_{AA}$ and $\tilde{C}_{AP} = \sqrt{\tilde{C}_{AA}\tilde{C}_{PP}}$.

Next we estimate the amount of classical noise in the laser, using two independent measurements. Based upon measurements in \cite{Peterson(16)_BALcool} we put an upper bound on classical laser phase noise of $\tilde{C}_{PP}=4\%$ and amplitude noise of $\tilde{C}_{AA}=0.4\%$ at $5\:\unit{\mu W}$. Because of the choice of normalization there is a factor of two between these values and the reported values in Ref.~\cite{Peterson(16)_BALcool}.

For a second, better, estimation of the phase noise we use the light PSD far off-resonance to find the difference in shot-noise level. We do this by looking at the deviation of the fit off-set from the well known shot-noise level. By comparing this to the expected rise in shot-noise, for an on-resonance probe
\begin{align}
S_{\rm{LN}}(\omega,\phi) = 8\epsilon \left(\frac{\kappa}{2}\right)^2 \abs{\chi_c(\omega)}^2 \left(\tilde{C}_{AA} \cos^2\phi + \tilde{C}_{PP} \sin^2\phi - \tilde{C}_{AP} \sin2\phi \right)
\end{align}
From the $\phi=90^{\circ}$ data and using our cavity parameters we find $S_{\rm{LN}}(\omega,\phi=90^{\circ})=1.06 \tilde{C}_{PP}\le 0.015$, which gives $\tilde{C}_{PP}\le 1.5\:\%$. With this we estimate $\tilde{C}_{AP} = 0.78\:\%$.

From the above estimations we find that the total classical noise in our experiment was less than $1\:\%$. Moreover if we plug in our experimental parameters to find the cavity and mechanical susceptibility response far off resonance ($\rho=10$) we find that the effect of the classical noise is suppressed by an additional order of magnitude.  For this reason in the main text, and the analysis procedure, we have set the classical noise contribution to zero.

\section{The Force Power Spectral density} \label{sec: force sens}

For a full context of backaction effects presented in the Appendix calculations we present here the conversion between displacement spectral density and the force spectral density in both our dimensionless and also force units.  Any external force applied on the mechanical resonator will be shaped by the dimensionless mechanical susceptibility function $\abs{\tilde{\chi}_m(\omega)}^2$; and the dimensionless force PSD is given by,
\begin{align}
S_{ff}(\rho) = \abs{\tilde{\chi}_m(\rho)}^{-2} S_{xx}(\rho)
\end{align}
and the force SQL PSD value is given,
\begin{align}
S_{\rm{sql}}^f(\omega) = p_{\rm{zp}}^2\frac{1}{\abs{\chi_m(\omega)}} = p_{\rm{zp}}^2\frac{\Gamma}{2} \sqrt{1 + \rho^2}.
\end{align}
with units of newton square per hertz.  Here to achieve force units we used the ratio between the zero-point motion and the zero-point fluctuations ($p_{\rm{zp}} = \hbar/(2 x_{\rm{zp}}) [\unit{\frac{N}{Hz}}]$).

Then the PSD with real dimension is $S_{\rm{sql}}^f(\omega_m) S_{ff}(\rho)$. As can be seen, the best sensitivity is found on-resonance while off-resonance the sensitivity falls off. The optimal measurement phase (and power) for force is the same as for the position detection. Which means that the optimal PSD for force is given by
\begin{align}
S_{ff}(\rho,\phi_{opt},p) &=
2\left(n_{\rm{th}}+\frac{1}{2}\right) + \frac{1}{2 \epsilon p \abs{\tilde{\chi}_m(\rho)}^2 } + \frac{1}{2} p \left( 1 + (1-\epsilon)\rho^2 \right) \abs{\tilde{\chi}_m(\rho)}^2 \\
S_{ff}(\rho,\phi_{opt},p_{opt}) &=
2\left(n_{\rm{th}}+\frac{1}{2}\right) + \sqrt{ \frac{1}{\epsilon} + \frac{1-\epsilon}{\epsilon}\rho^2 }
\end{align}
and in the last line we write the result for optimal power, i.e. the QL value for each frequency.

\section{Synodyne detection}

\begin{figure}[b]
	\centering
    \includegraphics[trim={1.5cm 20.25cm 2.25cm 0.25cm},clip]{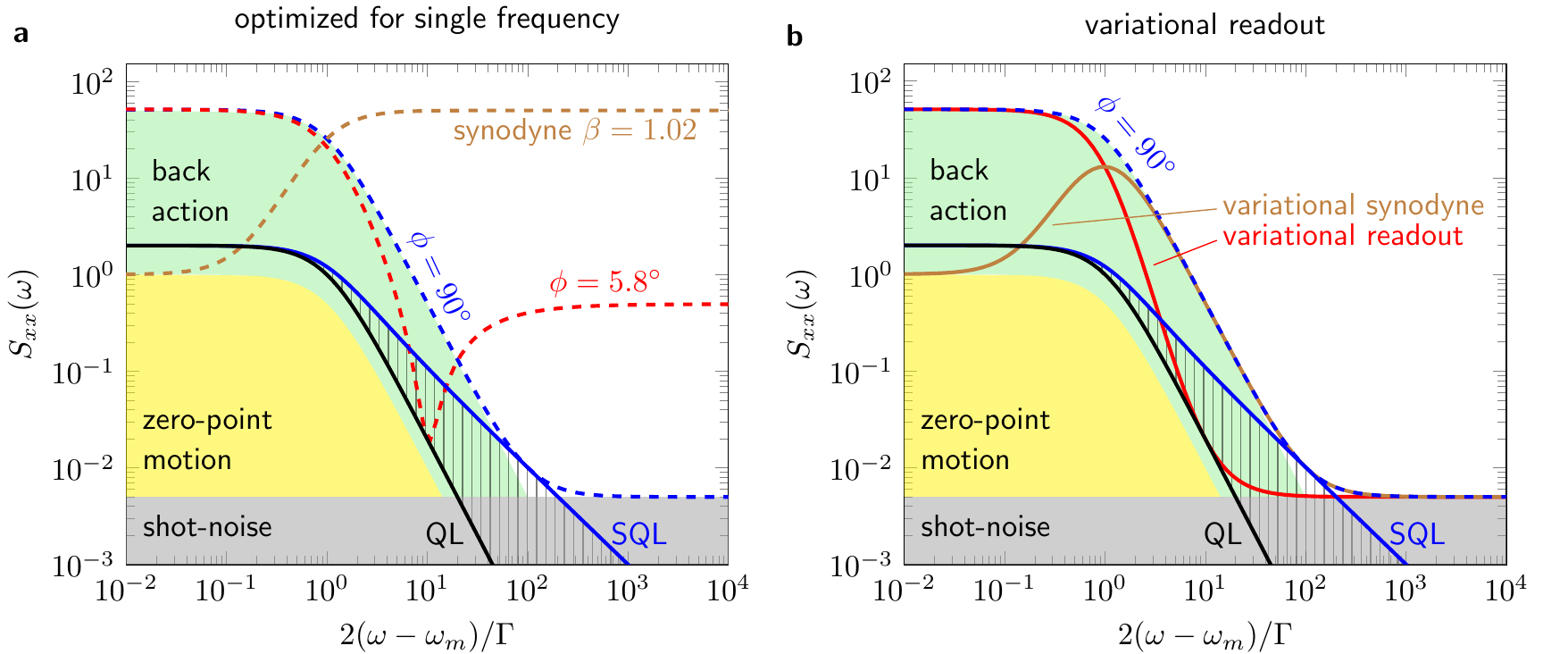}  
	\caption{Comparison of homodyne and synodyne detection at constant power ($p=100$). (a) Optimizing the spectrum for a single frequency, on-resonance for synodyne (brown) and $2(\omega-\omega_m)/\Gamma=10$ for homodyne (red); (b) variational readout in which we vary the local oscillator to optimize (minimize) the spectrum for all frequencies.  Shaded areas represent the shot-noise (gray), backaction (green) and zero-point motion (yellow). Lines represent measurement at $\phi=90^{\circ}$ (blue), $\phi=5.8^{\circ}$ (solid red), ratio of $\beta=1.02$ (solid brown), homodyne variational readout (red dashed), synodyne variational readout (brown dashed), SQL noise with zero thermal disturbance (solid black), QL noise with zero thermal disturbance (dashed black). All curves are evaluated for ideal quantum efficiency $\epsilon=1$, and zero thermal disturbance.  }
	\label{fig: supp spec}
\end{figure}

In the recently-proposed synodyne measurement~\cite{Buchmann(16)_Synodyne_Supp}, the cross-correlation discussed can be accessed on-resonance, which enables beating both the on-resonance SQL and QL.  Up until recently it was believed that only a quantum non-demolition (QND) measurement configuration~\cite{Clerk(08)_BAEsqueezingSupp,Braginsky(80)_QNDdefSupp} can improve on-resonance sensitivity below the SQL or QL limit.  This is allowed because in a QND measurement only a single mechanical quadrature is measured and one could reach a total noise of solely the zero point motion (yellow in Fig.~\ref{fig: supp spec}).  Synodyne similarly allows one to gain information about only one mechanical quadrature when measuring on mechanical resonance.

To detect only a single mechanical quadrature one can change the readout configuration even beyond that of variational homodyne by changing the character of the local oscillator (LO).  In particular one uses a two-tone LO with the tones split by twice the mechanical resonance frequency ($2\omega_m$).  This means that the signal is split into lower and upper sidebands, and the interference between the two sidebands effectively shifts the mechanical resonator to DC ($\omega_m\rightarrow 0$); which corresponds to a single mechanical-quadrature measurement.
After arranging the two LO tones, in analogy to variational homodyne, it is also necessary to arrange the contribution of each tone (magnitude and phase) such that the correlations will destructively interfere to remove the backaction contribution.  When the two sidebands are equal then the correlation contribution falls out, and the optimal measurement is similar to homodyne measurement at $\phi=90^{\circ}$.  But by using slightly different sidebands amplitudes, one mixes the probe's two quadratures and the on-resonance correlation contribution improves the measurement below the QL.  In Fig.~\ref{fig: supp spec} we show the complementary effects of synodyne and homodyne via their frequency-dependent noise PSD.

The purpose of the rest of this section is to give the basis for comparison between synodyne measurement configuration and homodyne measurement configuration in a frequency-space picture; and present a similar plot to Fig.~1(a,b) in the main text.
We start by explicitly writing the two-tone LO $\alpha(t) = \alpha_- e^{i\omega_m t} + \alpha_+ e^{-i\omega_m t}$, and define its amplitude and phase contribution as:
\begin{align}
\alpha_a =  \frac{1}{2}\left[ \alpha_- e^{-i\phi} + \alpha_+ e^{i\phi} \right] = \frac{\alpha_-}{2}\left[ e^{-i\phi} + \beta e^{i\phi} \right] \\
\alpha_p = -\frac{i}{2}\left[ \alpha_- e^{-i\phi} - \alpha_+ e^{i\phi} \right] = -\frac{i\alpha_-}{2}\left[ e^{-i\phi} - \beta e^{i\phi} \right]
\end{align}
where $\alpha_{\pm}$ are real, and $\beta=\alpha_+/\alpha_-$.  We note that in general one can add an additional phase between $\alpha_-$ and $\alpha_+$, but that only modifies the global phase $\phi$.  With this definition of the LO, we write a synodyne optical operator in the time domain as $\left[ \alpha_a \hat{u}_{AM}(t) + \alpha_p \hat{u}_{PM}(t) \right] e^{-i\omega_m t} + \left[ \alpha_a^* \hat{u}_{AM}(t) + \alpha_p^* \hat{u}_{PM}(t) \right] e^{i\omega_m t}$, where $\hat{u}_{AM}$ and $\hat{u}_{PM}$ are defined in Sec.~\ref{sec: light psd} in Eq.~\ref{eq: u mat} (after moving to the frequency domain).

Using this definition of LO, we find a new displacement operator. Note that in the mechanical susceptibility the mechanical resonance frequency is now $\omega_m=0$.  The spectrum is written similarly to homodyne detection. But, now unlike in the homodyne detection, the cross-correlation contribution is complex and $S_{IF} = -S_{FI}^*$.

The PSD for synodyne detection is given by
\begin{align}
S_{xx}^S(\rho,p) = 2\left(n_{\rm{th}}+\frac{1}{2}\right) \abs{\tilde{\chi}_{m}(\rho)}^2 + \frac{1}{2 \epsilon p}\frac{\abs{\alpha_a}^2 + \abs{\alpha_p}^2 }{\abs{\alpha_p}^2} + \frac{p}{2} \abs{\tilde{\chi}_m(\rho)}^2  - \abs{\tilde{\chi}_m(\rho)}^2 \Im \frac{ \alpha_a^*\alpha_p }{\abs{\alpha_p}^2} \label{eq: syno Sxx plain}
\end{align}
with
\begin{align}
\frac{\abs{\alpha_a}^2 + \abs{\alpha_p}^2}{\abs{\alpha_p}^2} =  2\frac{ 1 + \beta^2 }{ 1 + \beta^2 - 2\beta \cos2\phi } \\
\Im\frac{\alpha_a^*\alpha_p}{\abs{\alpha_p}^2} = \frac{\beta^2 - 1 } {1 + \beta^2 - 2\beta \cos2\phi}
\end{align}
In the above we have assumed that the cavity susceptibility's relative change is negligible, i.e. $\abs{\tilde{\chi}_c(\omega)}^2 \approx 1$. Note that because here we split the signal into positive and negative parts, this assumption also needs to hold as $\abs{\tilde{\chi}_c(\omega + \delta)}^2 \approx \abs{\tilde{\chi}_c(\omega - \delta)}^2$.

First we examine the case in which the two sidebands of the LO are equal, i.e. $\beta=1$. In this case the cross-correlation contribution falls out, and we get back the same results as in homodyne detection (without the cross-correlation term). Note the shot-noise term is proportional to $1+\cot^2\phi$, similar to the homodyne detection.

Next we find the optimal conditions for the phase between the two LO tones and the signal beam ($\phi$) and the ratio between the two LO tones $\beta$. Numerically we find that for low power it's best to set $\phi = 90^{\circ}$ and for powers above $p> 1/\epsilon \abs{\tilde{\chi}_m(\rho)}^2$ we should set $\phi = 0^{\circ}$. Then we find the optimal ratio is given by,
\begin{align}
\beta_{\rm{opt}}(\rho,\phi=90^{\circ}) &= \frac{1 + \epsilon p \abs{\tilde{\chi}_m(\rho)}^2}{1 - \epsilon p \abs{\tilde{\chi}_m(\rho)}^2} \\
\beta_{\rm{opt}}(\rho,\phi=0^{\circ}) &= \frac{\epsilon p \abs{\tilde{\chi}_m(\rho)}^2 + 1}{\epsilon p \abs{\tilde{\chi}_m(\rho)}^2 - 1} = -\beta_{\rm{opt}}(\rho,\phi=90^{\circ})
\end{align}

In Fig.~\ref{fig: supp spec}(a) we show the result of Eq.~(\ref{eq: syno Sxx plain}) (using a single value for $\beta$ optimal for on-resonance detection). The on-resonance PSD reaches the zero-point motion level, for ideal quantum efficiency. In addition the trade off between improvement at a given frequency comes at expense of additional shot-noise contribution (off-resonance), coming from the quadrature that does not include any position information.

Next we enter the optimal ratio at all frequencies, at constant power, to find the variational synodyne PSD,
\begin{align}
S^S_{xx}(\rho,p,\beta_{opt}(\rho)) =
2\left(n_{\rm{th}}+\frac{1}{2}\right) \abs{\tilde{\chi}_m(\rho)}^2 + \frac{1}{2\epsilon p} + \frac{1}{2}p \left[ (1 - \epsilon) + \rho^2\right] \abs{\tilde{\chi}_m(\rho)}^4 \label{eq: Sx-sql-opt syno}
\end{align}
This result is very similar to the homodyne variational readout, up to the effect of the quantum efficiency in the backaction term, i.e. $(1 - \epsilon) + \rho^2 \leftrightarrow 1 + (1 - \epsilon)\rho^2$.  Then the optimal synodyne power is,
\begin{align}
p_{\rm{opt}}^s = \frac{1}{\sqrt{ \epsilon \left[(1-\epsilon) +\rho^2\right]} \abs{\tilde{\chi}_m(\rho)}^2}
\end{align}
which gives,
\begin{align}
S^S_{xx}(\rho,p_{\rm{opt}}^s,\beta_{\rm{opt}}) = 2\left(n_{\rm{th}}+\frac{1}{2}\right) \abs{\tilde{\chi}_m(\rho)}^2 + \sqrt{ \frac{1-\epsilon}{\epsilon} + \frac{1}{\epsilon}\rho^2 } \abs{\tilde{\chi}_m(\rho)}^2
\end{align}
Here we find, as expected, that on-resonance the minimum PSD is equal (or bigger) to one zero-point motion. As we expect from a single mechanical quadrature measurement.
While there is a strong resemblance to a QND measurement, there are two differences: the frequency response, and the quantum efficiency dependence.

In Fig.~\ref{fig: supp spec}(b) the frequency response of variational synodyne readout is given. While at frequencies $\rho<1$ the synodyne variational readout gives the best result; at off-resonance frequencies ($\rho\ge1$) homodyne variational readout gives the best result. This result is expected because off-resonance synodyne detection is reduced to a homodyne measurement at $\phi=90^{\circ}$, i.e. limited by the SQL.

Because force sensitivity is highest on resonance (see Sec.~\ref{sec: force sens}), synodyne detection is especially good for force detection, with $S_{xx}^S/\abs{\tilde{\chi}_m(\rho)}^2$.  Therefore we explicitly write the displacement PSD due to external force modulated at $\omega_f$ to be
\begin{align}
\left(\frac{F}{4p_{\rm{zp}}}\right)^2 \abs{\tilde{\chi}_m(\rho)}^2 \times 
\frac{ \abs{ \alpha_p e^{-i\phi_f} \delta(\omega-\omega_f+\omega_m) + \alpha_p^* e^{i\phi_f} \delta(\omega+\omega_f-\omega_m) }^2 }{2\abs{\alpha_p}^2 }
\end{align}
with $\phi_f$ the force relative phase.
The above force PSD measures a single phase only for an on-resonance force ($\omega_f=\omega_m$), and for an off-mechanical-resonance the force is phase independent.  This appears because in synodyne detection, the measurement is shifted down to DC, at which the force phase is important.  This result resembles a BAE type measurement, but when examining the signal (both force sensitivity and noise PSD) there is a difference.


\begin{thebibliography}{46}%
\makeatletter
\providecommand \@ifxundefined [1]{%
 \@ifx{#1\undefined}
}%
\providecommand \@ifnum [1]{%
 \ifnum #1\expandafter \@firstoftwo
 \else \expandafter \@secondoftwo
 \fi
}%
\providecommand \@ifx [1]{%
 \ifx #1\expandafter \@firstoftwo
 \else \expandafter \@secondoftwo
 \fi
}%
\providecommand \natexlab [1]{#1}%
\providecommand \enquote  [1]{``#1''}%
\providecommand \bibnamefont  [1]{#1}%
\providecommand \bibfnamefont [1]{#1}%
\providecommand \citenamefont [1]{#1}%
\providecommand \href@noop [0]{\@secondoftwo}%
\providecommand \href [0]{\begingroup \@sanitize@url \@href}%
\providecommand \@href[1]{\@@startlink{#1}\@@href}%
\providecommand \@@href[1]{\endgroup#1\@@endlink}%
\providecommand \@sanitize@url [0]{\catcode `\\12\catcode `\$12\catcode
  `\&12\catcode `\#12\catcode `\^12\catcode `\_12\catcode `\%12\relax}%
\providecommand \@@startlink[1]{}%
\providecommand \@@endlink[0]{}%
\providecommand \url  [0]{\begingroup\@sanitize@url \@url }%
\providecommand \@url [1]{\endgroup\@href {#1}{\urlprefix }}%
\providecommand \urlprefix  [0]{URL }%
\providecommand \Eprint [0]{\href }%
\providecommand \doibase [0]{http://dx.doi.org/}%
\providecommand \selectlanguage [0]{\@gobble}%
\providecommand \bibinfo  [0]{\@secondoftwo}%
\providecommand \bibfield  [0]{\@secondoftwo}%
\providecommand \translation [1]{[#1]}%
\providecommand \BibitemOpen [0]{}%
\providecommand \bibitemStop [0]{}%
\providecommand \bibitemNoStop [0]{.\EOS\space}%
\providecommand \EOS [0]{\spacefactor3000\relax}%
\providecommand \BibitemShut  [1]{\csname bibitem#1\endcsname}%
\let\auto@bib@innerbib\@empty
\bibitem [{\citenamefont {Haus}\ and\ \citenamefont {Mullen}(1962)}]{Haus62}%
  \BibitemOpen
  \bibfield  {author} {\bibinfo {author} {\bibfnamefont {H.~A.}\ \bibnamefont
  {Haus}}\ and\ \bibinfo {author} {\bibfnamefont {J.~A.}\ \bibnamefont
  {Mullen}},\ }\bibfield  {title} {\enquote {\bibinfo {title} {Quantum noise in
  linear amplifiers},}\ }\href {\doibase 10.1103/PhysRev.128.2407} {\bibfield
  {journal} {\bibinfo  {journal} {Phys. Rev.}\ }\textbf {\bibinfo {volume}
  {128}},\ \bibinfo {pages} {2407--2413} (\bibinfo {year} {1962})}\BibitemShut
  {NoStop}%
\bibitem [{\citenamefont {Caves}(1982)}]{Caves82}%
  \BibitemOpen
  \bibfield  {author} {\bibinfo {author} {\bibfnamefont {Carlton~M.}\
  \bibnamefont {Caves}},\ }\bibfield  {title} {\enquote {\bibinfo {title}
  {Quantum limits on noise in linear amplifiers},}\ }\href {\doibase
  10.1103/PhysRevD.26.1817} {\bibfield  {journal} {\bibinfo  {journal} {Phys.
  Rev. D}\ }\textbf {\bibinfo {volume} {26}},\ \bibinfo {pages} {1817--1839}
  (\bibinfo {year} {1982})}\BibitemShut {NoStop}%
\bibitem [{\citenamefont {Braginsky}\ \emph {et~al.}(2003)\citenamefont
  {Braginsky}, \citenamefont {Gorodetsky}, \citenamefont {Khalili},
  \citenamefont {Matsko}, \citenamefont {Thorne},\ and\ \citenamefont
  {Vyatchanin}}]{Braginsky(03)_NonRelaventSQL}%
  \BibitemOpen
  \bibfield  {author} {\bibinfo {author} {\bibfnamefont {Vladimir~B.}\
  \bibnamefont {Braginsky}}, \bibinfo {author} {\bibfnamefont {Mikhail~L.}\
  \bibnamefont {Gorodetsky}}, \bibinfo {author} {\bibfnamefont {Farid~Ya.}\
  \bibnamefont {Khalili}}, \bibinfo {author} {\bibfnamefont {Andrey~B.}\
  \bibnamefont {Matsko}}, \bibinfo {author} {\bibfnamefont {Kip~S.}\
  \bibnamefont {Thorne}}, \ and\ \bibinfo {author} {\bibfnamefont {Sergey~P.}\
  \bibnamefont {Vyatchanin}},\ }\bibfield  {title} {\enquote {\bibinfo {title}
  {Noise in gravitational-wave detectors and other classical-force measurements
  is not influenced by test-mass quantization},}\ }\href {\doibase
  10.1103/PhysRevD.67.082001} {\bibfield  {journal} {\bibinfo  {journal} {Phys.
  Rev. D}\ }\textbf {\bibinfo {volume} {67}},\ \bibinfo {pages} {082001}
  (\bibinfo {year} {2003})}\BibitemShut {NoStop}%
\bibitem [{\citenamefont {Clerk}\ \emph {et~al.}(2010)\citenamefont {Clerk},
  \citenamefont {Devoret}, \citenamefont {Girvin}, \citenamefont {Marquardt},\
  and\ \citenamefont {Schoelkopf}}]{Clerk(10)_QN}%
  \BibitemOpen
  \bibfield  {author} {\bibinfo {author} {\bibfnamefont {A.~A.}\ \bibnamefont
  {Clerk}}, \bibinfo {author} {\bibfnamefont {M.~H.}\ \bibnamefont {Devoret}},
  \bibinfo {author} {\bibfnamefont {S.~M.}\ \bibnamefont {Girvin}}, \bibinfo
  {author} {\bibfnamefont {Florian}\ \bibnamefont {Marquardt}}, \ and\ \bibinfo
  {author} {\bibfnamefont {R.~J.}\ \bibnamefont {Schoelkopf}},\ }\bibfield
  {title} {\enquote {\bibinfo {title} {Introduction to quantum noise,
  measurement, and amplification},}\ }\href {\doibase
  10.1103/RevModPhys.82.1155} {\bibfield  {journal} {\bibinfo  {journal} {Rev.
  Mod. Phys.}\ }\textbf {\bibinfo {volume} {82}},\ \bibinfo {pages}
  {1155--1208} (\bibinfo {year} {2010})}\BibitemShut {NoStop}%
\bibitem [{\citenamefont {Jaekel}\ and\ \citenamefont
  {Reynaud}(1990)}]{Jaekel(90)_QL}%
  \BibitemOpen
  \bibfield  {author} {\bibinfo {author} {\bibfnamefont {M.~T.}\ \bibnamefont
  {Jaekel}}\ and\ \bibinfo {author} {\bibfnamefont {S.}~\bibnamefont
  {Reynaud}},\ }\bibfield  {title} {\enquote {\bibinfo {title} {Quantum limits
  in interferometric measurements},}\ }\href
  {http://stacks.iop.org/0295-5075/13/i=4/a=003} {\bibfield  {journal}
  {\bibinfo  {journal} {Europhys. Lett.}\ }\textbf {\bibinfo {volume} {13}},\
  \bibinfo {pages} {301} (\bibinfo {year} {1990})}\BibitemShut {NoStop}%
\bibitem [{\citenamefont {Braginskii}\ and\ \citenamefont
  {Vorontsov}(1968)}]{Braginskii(68)_SQLfirst}%
  \BibitemOpen
  \bibfield  {author} {\bibinfo {author} {\bibfnamefont {V.~B.}\ \bibnamefont
  {Braginskii}}\ and\ \bibinfo {author} {\bibfnamefont {Yu.~I.}\ \bibnamefont
  {Vorontsov}},\ }\bibfield  {title} {\enquote {\bibinfo {title} {Classical and
  quantum restrictions on the detection of weak disturbances of a macroscopic
  oscillator},}\ }\href
  {http://www.jetp.ac.ru/cgi-bin/e/index/e/26/4/p831?a=list} {\bibfield
  {journal} {\bibinfo  {journal} {Sov. Phys. JETP}\ }\textbf {\bibinfo {volume}
  {26}},\ \bibinfo {pages} {831--834} (\bibinfo {year} {1968})}\BibitemShut
  {NoStop}%
\bibitem [{\citenamefont {Caves}(1980)}]{Caves(80)_QM_RPinInter}%
  \BibitemOpen
  \bibfield  {author} {\bibinfo {author} {\bibfnamefont {Carlton~M.}\
  \bibnamefont {Caves}},\ }\bibfield  {title} {\enquote {\bibinfo {title}
  {Quantum-mechanical radiation-pressure fluctuations in an interferometer},}\
  }\href {\doibase 10.1103/PhysRevLett.45.75} {\bibfield  {journal} {\bibinfo
  {journal} {Phys. Rev. Lett.}\ }\textbf {\bibinfo {volume} {45}},\ \bibinfo
  {pages} {75--79} (\bibinfo {year} {1980})}\BibitemShut {NoStop}%
\bibitem [{\citenamefont {LaHaye}\ \emph {et~al.}(2004)\citenamefont {LaHaye},
  \citenamefont {Buu}, \citenamefont {Camarota},\ and\ \citenamefont
  {Schwab}}]{LaHaye(04)_appQL}%
  \BibitemOpen
  \bibfield  {author} {\bibinfo {author} {\bibfnamefont {M.~D.}\ \bibnamefont
  {LaHaye}}, \bibinfo {author} {\bibfnamefont {O.}~\bibnamefont {Buu}},
  \bibinfo {author} {\bibfnamefont {B.}~\bibnamefont {Camarota}}, \ and\
  \bibinfo {author} {\bibfnamefont {K.~C.}\ \bibnamefont {Schwab}},\ }\bibfield
   {title} {\enquote {\bibinfo {title} {Approaching the quantum limit of a
  nanomechanical resonator},}\ }\href {\doibase 10.1126/science.1094419}
  {\bibfield  {journal} {\bibinfo  {journal} {Science}\ }\textbf {\bibinfo
  {volume} {304}},\ \bibinfo {pages} {74--77} (\bibinfo {year}
  {2004})}\BibitemShut {NoStop}%
\bibitem [{\citenamefont {Purdy}\ \emph {et~al.}(2012)\citenamefont {Purdy},
  \citenamefont {Peterson},\ and\ \citenamefont {Regal}}]{Purdy(12)_RPSN}%
  \BibitemOpen
  \bibfield  {author} {\bibinfo {author} {\bibfnamefont {T.~P.}\ \bibnamefont
  {Purdy}}, \bibinfo {author} {\bibfnamefont {R.~W.}\ \bibnamefont {Peterson}},
  \ and\ \bibinfo {author} {\bibfnamefont {C.~A.}\ \bibnamefont {Regal}},\
  }\bibfield  {title} {\enquote {\bibinfo {title} {Observation of radiation
  pressure shot noise on a macroscopic object},}\ }\href {\doibase
  10.1126/science.1231282} {\bibfield  {journal} {\bibinfo  {journal}
  {Science}\ }\textbf {\bibinfo {volume} {339}},\ \bibinfo {pages} {801--804}
  (\bibinfo {year} {2012})}\BibitemShut {NoStop}%
\bibitem [{\citenamefont {Safavi-Naeini}\ \emph {et~al.}(2012)\citenamefont
  {Safavi-Naeini}, \citenamefont {Chan}, \citenamefont {Hill}, \citenamefont
  {Alegre}, \citenamefont {Krause},\ and\ \citenamefont
  {Painter}}]{SafaviNaeini(12)_RPSN}%
  \BibitemOpen
  \bibfield  {author} {\bibinfo {author} {\bibfnamefont {Amir~H.}\ \bibnamefont
  {Safavi-Naeini}}, \bibinfo {author} {\bibfnamefont {Jasper}\ \bibnamefont
  {Chan}}, \bibinfo {author} {\bibfnamefont {Jeff~T.}\ \bibnamefont {Hill}},
  \bibinfo {author} {\bibfnamefont {Thiago P.~Mayer}\ \bibnamefont {Alegre}},
  \bibinfo {author} {\bibfnamefont {Alex}\ \bibnamefont {Krause}}, \ and\
  \bibinfo {author} {\bibfnamefont {Oskar}\ \bibnamefont {Painter}},\
  }\bibfield  {title} {\enquote {\bibinfo {title} {Observation of quantum
  motion of a nanomechanical resonator},}\ }\href {\doibase
  10.1103/PhysRevLett.108.033602} {\bibfield  {journal} {\bibinfo  {journal}
  {Phys. Rev. Lett.}\ }\textbf {\bibinfo {volume} {108}},\ \bibinfo {pages}
  {033602} (\bibinfo {year} {2012})}\BibitemShut {NoStop}%
\bibitem [{\citenamefont {Schreppler}\ \emph {et~al.}(2014)\citenamefont
  {Schreppler}, \citenamefont {Spethmann}, \citenamefont {Brahms},
  \citenamefont {Botter}, \citenamefont {Barrios},\ and\ \citenamefont
  {Stamper-Kurn}}]{Schreppler(14)_SQL}%
  \BibitemOpen
  \bibfield  {author} {\bibinfo {author} {\bibfnamefont {Sydney}\ \bibnamefont
  {Schreppler}}, \bibinfo {author} {\bibfnamefont {Nicolas}\ \bibnamefont
  {Spethmann}}, \bibinfo {author} {\bibfnamefont {Nathan}\ \bibnamefont
  {Brahms}}, \bibinfo {author} {\bibfnamefont {Thierry}\ \bibnamefont
  {Botter}}, \bibinfo {author} {\bibfnamefont {Maryrose}\ \bibnamefont
  {Barrios}}, \ and\ \bibinfo {author} {\bibfnamefont {Dan~M.}\ \bibnamefont
  {Stamper-Kurn}},\ }\bibfield  {title} {\enquote {\bibinfo {title} {Optically
  measuring force near the standard quantum limit},}\ }\href {\doibase
  10.1126/science.1249850} {\bibfield  {journal} {\bibinfo  {journal}
  {Science}\ }\textbf {\bibinfo {volume} {344}},\ \bibinfo {pages} {1486--1489}
  (\bibinfo {year} {2014})}\BibitemShut {NoStop}%
\bibitem [{\citenamefont {Teufel}\ \emph {et~al.}(2016)\citenamefont {Teufel},
  \citenamefont {Lecocq},\ and\ \citenamefont
  {Simmonds}}]{Teufel(16)_StrongRPSN}%
  \BibitemOpen
  \bibfield  {author} {\bibinfo {author} {\bibfnamefont {J.~D.}\ \bibnamefont
  {Teufel}}, \bibinfo {author} {\bibfnamefont {F.}~\bibnamefont {Lecocq}}, \
  and\ \bibinfo {author} {\bibfnamefont {R.~W.}\ \bibnamefont {Simmonds}},\
  }\bibfield  {title} {\enquote {\bibinfo {title} {Overwhelming
  thermomechanical motion with microwave radiation pressure shot noise},}\
  }\href {\doibase 10.1103/PhysRevLett.116.013602} {\bibfield  {journal}
  {\bibinfo  {journal} {Phys. Rev. Lett.}\ }\textbf {\bibinfo {volume} {116}},\
  \bibinfo {pages} {013602} (\bibinfo {year} {2016})}\BibitemShut {NoStop}%
\bibitem [{\citenamefont {Miao}\ \emph {et~al.}(2014)\citenamefont {Miao},
  \citenamefont {Yang}, \citenamefont {Adhikari},\ and\ \citenamefont
  {Chen}}]{Miao(14)_Interef_lim_GW}%
  \BibitemOpen
  \bibfield  {author} {\bibinfo {author} {\bibfnamefont {Haixing}\ \bibnamefont
  {Miao}}, \bibinfo {author} {\bibfnamefont {Huan}\ \bibnamefont {Yang}},
  \bibinfo {author} {\bibfnamefont {Rana~X}\ \bibnamefont {Adhikari}}, \ and\
  \bibinfo {author} {\bibfnamefont {Yanbei}\ \bibnamefont {Chen}},\ }\bibfield
  {title} {\enquote {\bibinfo {title} {Quantum limits of interferometer
  topologies for gravitational radiation detection},}\ }\href {\doibase
  10.1088/0264-9381/31/16/165010} {\bibfield  {journal} {\bibinfo  {journal}
  {Classical and Quantum Gravity}\ }\textbf {\bibinfo {volume} {31}},\ \bibinfo
  {pages} {165010} (\bibinfo {year} {2014})}\BibitemShut {NoStop}%
\bibitem [{\citenamefont {Vyatchanin}\ and\ \citenamefont
  {Zubova}(1995)}]{Vyatchanin}%
  \BibitemOpen
  \bibfield  {author} {\bibinfo {author} {\bibfnamefont {S.~P.}\ \bibnamefont
  {Vyatchanin}}\ and\ \bibinfo {author} {\bibfnamefont {E.~A.}\ \bibnamefont
  {Zubova}},\ }\bibfield  {title} {\enquote {\bibinfo {title} {Quantum
  variation measurement of a force},}\ }\href {\doibase
  10.1016/0375-9601(95)00280-G} {\bibfield  {journal} {\bibinfo  {journal}
  {Physics Letters A}\ }\textbf {\bibinfo {volume} {201}},\ \bibinfo {pages}
  {269} (\bibinfo {year} {1995})}\BibitemShut {NoStop}%
\bibitem [{\citenamefont {Kimble}\ \emph {et~al.}(2001)\citenamefont {Kimble},
  \citenamefont {Levin}, \citenamefont {Matsko}, \citenamefont {Thorne},\ and\
  \citenamefont {Vyatchanin}}]{Kimble(01)_GI_inter}%
  \BibitemOpen
  \bibfield  {author} {\bibinfo {author} {\bibfnamefont {H.~J.}\ \bibnamefont
  {Kimble}}, \bibinfo {author} {\bibfnamefont {Yuri}\ \bibnamefont {Levin}},
  \bibinfo {author} {\bibfnamefont {Andrey~B.}\ \bibnamefont {Matsko}},
  \bibinfo {author} {\bibfnamefont {Kip~S.}\ \bibnamefont {Thorne}}, \ and\
  \bibinfo {author} {\bibfnamefont {Sergey~P.}\ \bibnamefont {Vyatchanin}},\
  }\bibfield  {title} {\enquote {\bibinfo {title} {Conversion of conventional
  gravitational-wave interferometers into quantum nondemolition interferometers
  by modifying their input and/or output optics},}\ }\href {\doibase
  10.1103/PhysRevD.65.022002} {\bibfield  {journal} {\bibinfo  {journal} {Phys.
  Rev. D}\ }\textbf {\bibinfo {volume} {65}},\ \bibinfo {pages} {022002}
  (\bibinfo {year} {2001})}\BibitemShut {NoStop}%
\bibitem [{\citenamefont {Braginsky}\ \emph {et~al.}(1980)\citenamefont
  {Braginsky}, \citenamefont {Vorontsov},\ and\ \citenamefont
  {Thorne}}]{Braginsky(80)_QNDdef}%
  \BibitemOpen
  \bibfield  {author} {\bibinfo {author} {\bibfnamefont {Vladimir~B.}\
  \bibnamefont {Braginsky}}, \bibinfo {author} {\bibfnamefont {Yuri~I.}\
  \bibnamefont {Vorontsov}}, \ and\ \bibinfo {author} {\bibfnamefont {Kip~S.}\
  \bibnamefont {Thorne}},\ }\bibfield  {title} {\enquote {\bibinfo {title}
  {Quantum nondemolition measurements},}\ }\href {\doibase
  10.1126/science.209.4456.547} {\bibfield  {journal} {\bibinfo  {journal}
  {Science}\ }\textbf {\bibinfo {volume} {209}},\ \bibinfo {pages} {547--557}
  (\bibinfo {year} {1980})}\BibitemShut {NoStop}%
\bibitem [{\citenamefont {Hammerer}\ \emph {et~al.}(2009)\citenamefont
  {Hammerer}, \citenamefont {Aspelmeyer}, \citenamefont {Polzik},\ and\
  \citenamefont {Zoller}}]{Hammerer(09)_EPRchannelsMechAtom}%
  \BibitemOpen
  \bibfield  {author} {\bibinfo {author} {\bibfnamefont {K.}~\bibnamefont
  {Hammerer}}, \bibinfo {author} {\bibfnamefont {M.}~\bibnamefont
  {Aspelmeyer}}, \bibinfo {author} {\bibfnamefont {E.~S.}\ \bibnamefont
  {Polzik}}, \ and\ \bibinfo {author} {\bibfnamefont {P.}~\bibnamefont
  {Zoller}},\ }\bibfield  {title} {\enquote {\bibinfo {title} {Establishing
  einstein-poldosky-rosen channels between nanomechanics and atomic
  ensembles},}\ }\href {\doibase 10.1103/PhysRevLett.102.020501} {\bibfield
  {journal} {\bibinfo  {journal} {Phys. Rev. Lett.}\ }\textbf {\bibinfo
  {volume} {102}},\ \bibinfo {pages} {020501} (\bibinfo {year}
  {2009})}\BibitemShut {NoStop}%
\bibitem [{\citenamefont {Tsang}\ and\ \citenamefont
  {Caves}(2012)}]{Caves(12)_EvadQMmanyQuad}%
  \BibitemOpen
  \bibfield  {author} {\bibinfo {author} {\bibfnamefont {Mankei}\ \bibnamefont
  {Tsang}}\ and\ \bibinfo {author} {\bibfnamefont {Carlton~M.}\ \bibnamefont
  {Caves}},\ }\bibfield  {title} {\enquote {\bibinfo {title} {Evading quantum
  mechanics: Engineering a classical subsystem within a quantum environment},}\
  }\href {\doibase 10.1103/PhysRevX.2.031016} {\bibfield  {journal} {\bibinfo
  {journal} {Phys. Rev. X}\ }\textbf {\bibinfo {volume} {2}},\ \bibinfo {pages}
  {031016} (\bibinfo {year} {2012})}\BibitemShut {NoStop}%
\bibitem [{\citenamefont {Ockeloen-Korppia}\ \emph {et~al.}(2016)\citenamefont
  {Ockeloen-Korppia}, \citenamefont {Damskagg}, \citenamefont {Pirkkalainen},
  \citenamefont {Clerk}, \citenamefont {Woolley},\ and\ \citenamefont
  {Sillanpaa}}]{Ockeloen-Korppia2016}%
  \BibitemOpen
  \bibfield  {author} {\bibinfo {author} {\bibfnamefont {C.~F.}\ \bibnamefont
  {Ockeloen-Korppia}}, \bibinfo {author} {\bibfnamefont {E.}~\bibnamefont
  {Damskagg}}, \bibinfo {author} {\bibfnamefont {J.-M.}\ \bibnamefont
  {Pirkkalainen}}, \bibinfo {author} {\bibfnamefont {A.~A.}\ \bibnamefont
  {Clerk}}, \bibinfo {author} {\bibfnamefont {M.~J.}\ \bibnamefont {Woolley}},
  \ and\ \bibinfo {author} {\bibfnamefont {M.~A.}\ \bibnamefont {Sillanpaa}},\
  }\bibfield  {title} {\enquote {\bibinfo {title} {Quantum backaction evading
  measurement of collective mechanical modes},}\ }\href
  {https://doi.org/10.1103/PhysRevLett.117.140401} {\bibfield  {journal}
  {\bibinfo  {journal} {Phys. Rev. Lett.}\ }\textbf {\bibinfo {volume} {117}},\
  \bibinfo {pages} {140401} (\bibinfo {year} {2016})}\BibitemShut {NoStop}%
\bibitem [{\citenamefont {Moller}\ \emph {et~al.}(2016)\citenamefont {Moller},
  \citenamefont {Thomas}, \citenamefont {Vasilakis}, \citenamefont {Zeuthen},
  \citenamefont {Tsaturyan}, \citenamefont {Jensen}, \citenamefont
  {Schliesser}, \citenamefont {Hammerer},\ and\ \citenamefont
  {Polzik}}]{Moller2016}%
  \BibitemOpen
  \bibfield  {author} {\bibinfo {author} {\bibfnamefont {Christoffer~B.}\
  \bibnamefont {Moller}}, \bibinfo {author} {\bibfnamefont {Rodrigo~A.}\
  \bibnamefont {Thomas}}, \bibinfo {author} {\bibfnamefont {Georgios}\
  \bibnamefont {Vasilakis}}, \bibinfo {author} {\bibfnamefont {Emil}\
  \bibnamefont {Zeuthen}}, \bibinfo {author} {\bibfnamefont {Yeghishe}\
  \bibnamefont {Tsaturyan}}, \bibinfo {author} {\bibfnamefont {Kasper}\
  \bibnamefont {Jensen}}, \bibinfo {author} {\bibfnamefont {Albert}\
  \bibnamefont {Schliesser}}, \bibinfo {author} {\bibfnamefont {Klemens}\
  \bibnamefont {Hammerer}}, \ and\ \bibinfo {author} {\bibfnamefont
  {Eugene~S.}\ \bibnamefont {Polzik}},\ }\bibfield  {title} {\enquote {\bibinfo
  {title} {Back action evading quantum measurement of motion in a negative mass
  reference frame},}\ }\href {https://arxiv.org/abs/1608.03613} {\bibfield
  {journal} {\bibinfo  {journal} {arXiv:1608.03613}\ } (\bibinfo {year}
  {2016})}\BibitemShut {NoStop}%
\bibitem [{\citenamefont {Suh}\ \emph {et~al.}(2014)\citenamefont {Suh},
  \citenamefont {Weinstein}, \citenamefont {Lei}, \citenamefont {Wollman},
  \citenamefont {Steinke}, \citenamefont {Meystre}, \citenamefont {Clerk},\
  and\ \citenamefont {Schwab}}]{Suh(14)_MechBAE}%
  \BibitemOpen
  \bibfield  {author} {\bibinfo {author} {\bibfnamefont {J.}~\bibnamefont
  {Suh}}, \bibinfo {author} {\bibfnamefont {A.~J.}\ \bibnamefont {Weinstein}},
  \bibinfo {author} {\bibfnamefont {C.~U.}\ \bibnamefont {Lei}}, \bibinfo
  {author} {\bibfnamefont {E.~E.}\ \bibnamefont {Wollman}}, \bibinfo {author}
  {\bibfnamefont {S.~K.}\ \bibnamefont {Steinke}}, \bibinfo {author}
  {\bibfnamefont {P.}~\bibnamefont {Meystre}}, \bibinfo {author} {\bibfnamefont
  {A.~A.}\ \bibnamefont {Clerk}}, \ and\ \bibinfo {author} {\bibfnamefont
  {K.~C.}\ \bibnamefont {Schwab}},\ }\bibfield  {title} {\enquote {\bibinfo
  {title} {Mechanically detecting and avoiding the quantum fluctuations of a
  microwave field},}\ }\href {\doibase 10.1126/science.1253258} {\bibfield
  {journal} {\bibinfo  {journal} {Science}\ }\textbf {\bibinfo {volume}
  {344}},\ \bibinfo {pages} {1262--1265} (\bibinfo {year} {2014})}\BibitemShut
  {NoStop}%
\bibitem [{\citenamefont {Wollman}\ \emph {et~al.}(2015)\citenamefont
  {Wollman}, \citenamefont {Lei}, \citenamefont {Weinstein}, \citenamefont
  {Suh}, \citenamefont {Kronwald}, \citenamefont {Marquardt}, \citenamefont
  {Clerk},\ and\ \citenamefont {Schwab}}]{Wollman(15)_MechSqueezing}%
  \BibitemOpen
  \bibfield  {author} {\bibinfo {author} {\bibfnamefont {E.~E.}\ \bibnamefont
  {Wollman}}, \bibinfo {author} {\bibfnamefont {C.~U.}\ \bibnamefont {Lei}},
  \bibinfo {author} {\bibfnamefont {A.~J.}\ \bibnamefont {Weinstein}}, \bibinfo
  {author} {\bibfnamefont {J.}~\bibnamefont {Suh}}, \bibinfo {author}
  {\bibfnamefont {A.}~\bibnamefont {Kronwald}}, \bibinfo {author}
  {\bibfnamefont {F.}~\bibnamefont {Marquardt}}, \bibinfo {author}
  {\bibfnamefont {A.~A.}\ \bibnamefont {Clerk}}, \ and\ \bibinfo {author}
  {\bibfnamefont {K.~C.}\ \bibnamefont {Schwab}},\ }\bibfield  {title}
  {\enquote {\bibinfo {title} {Quantum squeezing of motion in a mechanical
  resonator},}\ }\href {\doibase 10.1126/science.aac5138} {\bibfield  {journal}
  {\bibinfo  {journal} {Science}\ }\textbf {\bibinfo {volume} {349}},\ \bibinfo
  {pages} {952--955} (\bibinfo {year} {2015})}\BibitemShut {NoStop}%
\bibitem [{\citenamefont {Lecocq}\ \emph {et~al.}(2015)\citenamefont {Lecocq},
  \citenamefont {Clark}, \citenamefont {Simmonds}, \citenamefont {Aumentado},\
  and\ \citenamefont {Teufel}}]{Lecocq(15)_MechSqueezing}%
  \BibitemOpen
  \bibfield  {author} {\bibinfo {author} {\bibfnamefont {F.}~\bibnamefont
  {Lecocq}}, \bibinfo {author} {\bibfnamefont {J.~B.}\ \bibnamefont {Clark}},
  \bibinfo {author} {\bibfnamefont {R.~W.}\ \bibnamefont {Simmonds}}, \bibinfo
  {author} {\bibfnamefont {J.}~\bibnamefont {Aumentado}}, \ and\ \bibinfo
  {author} {\bibfnamefont {J.~D.}\ \bibnamefont {Teufel}},\ }\bibfield  {title}
  {\enquote {\bibinfo {title} {Quantum nondemolition measurement of a
  nonclassical state of a massive object},}\ }\href {\doibase
  10.1103/PhysRevX.5.041037} {\bibfield  {journal} {\bibinfo  {journal} {Phys.
  Rev. X}\ }\textbf {\bibinfo {volume} {5}},\ \bibinfo {pages} {041037}
  (\bibinfo {year} {2015})}\BibitemShut {NoStop}%
\bibitem [{\citenamefont {Pirkkalainen}\ \emph {et~al.}(2015)\citenamefont
  {Pirkkalainen}, \citenamefont {Damsk\"agg}, \citenamefont {Brandt},
  \citenamefont {Massel},\ and\ \citenamefont
  {Sillanp\"a\"a}}]{Pirkkalainen(15)_MechSqueezing}%
  \BibitemOpen
  \bibfield  {author} {\bibinfo {author} {\bibfnamefont {J.-M.}\ \bibnamefont
  {Pirkkalainen}}, \bibinfo {author} {\bibfnamefont {E.}~\bibnamefont
  {Damsk\"agg}}, \bibinfo {author} {\bibfnamefont {M.}~\bibnamefont {Brandt}},
  \bibinfo {author} {\bibfnamefont {F.}~\bibnamefont {Massel}}, \ and\ \bibinfo
  {author} {\bibfnamefont {M.~A.}\ \bibnamefont {Sillanp\"a\"a}},\ }\bibfield
  {title} {\enquote {\bibinfo {title} {Squeezing of quantum noise of motion in
  a micromechanical resonator},}\ }\href {\doibase
  10.1103/PhysRevLett.115.243601} {\bibfield  {journal} {\bibinfo  {journal}
  {Phys. Rev. Lett.}\ }\textbf {\bibinfo {volume} {115}},\ \bibinfo {pages}
  {243601} (\bibinfo {year} {2015})}\BibitemShut {NoStop}%
\bibitem [{\citenamefont {Rehbein}\ \emph {et~al.}(2005)\citenamefont
  {Rehbein}, \citenamefont {Harms}, \citenamefont {Schnabel},\ and\
  \citenamefont {Danzmann}}]{Rehbein2005}%
  \BibitemOpen
  \bibfield  {author} {\bibinfo {author} {\bibfnamefont {Henning}\ \bibnamefont
  {Rehbein}}, \bibinfo {author} {\bibfnamefont {Jan}\ \bibnamefont {Harms}},
  \bibinfo {author} {\bibfnamefont {Roman}\ \bibnamefont {Schnabel}}, \ and\
  \bibinfo {author} {\bibfnamefont {Karsten}\ \bibnamefont {Danzmann}},\
  }\bibfield  {title} {\enquote {\bibinfo {title} {Optical transfer functions
  of kerr nonlinear cavities and interferometers},}\ }\href
  {http://journals.aps.org/prl/abstract/10.1103/PhysRevLett.95.193001}
  {\bibfield  {journal} {\bibinfo  {journal} {Phys. Rev. Lett.}\ }\textbf
  {\bibinfo {volume} {95}},\ \bibinfo {pages} {193001} (\bibinfo {year}
  {2005})}\BibitemShut {NoStop}%
\bibitem [{\citenamefont {Laflamme}\ and\ \citenamefont
  {Clerk}(2011)}]{Laflamme2011}%
  \BibitemOpen
  \bibfield  {author} {\bibinfo {author} {\bibfnamefont {C.}~\bibnamefont
  {Laflamme}}\ and\ \bibinfo {author} {\bibfnamefont {A.~A.}\ \bibnamefont
  {Clerk}},\ }\bibfield  {title} {\enquote {\bibinfo {title} {Quantum-limited
  amplification with a nonlinear cavity detector},}\ }\href
  {http://journals.aps.org/pra/abstract/10.1103/PhysRevA.83.033803} {\bibfield
  {journal} {\bibinfo  {journal} {Phys. Rev. A}\ }\textbf {\bibinfo {volume}
  {83}},\ \bibinfo {pages} {033803} (\bibinfo {year} {2011})}\BibitemShut
  {NoStop}%
\bibitem [{\citenamefont {Braginsky}\ and\ \citenamefont
  {Khalili}(1992)}]{BraginskyKhalili(92)_book}%
  \BibitemOpen
  \bibfield  {author} {\bibinfo {author} {\bibfnamefont {Vladimir~B.}\
  \bibnamefont {Braginsky}}\ and\ \bibinfo {author} {\bibfnamefont {Farid~Ya.}\
  \bibnamefont {Khalili}},\ }\href@noop {} {\emph {\bibinfo {title} {Quantum
  Measurement}}}\ (\bibinfo  {publisher} {Cambridge University Press},\
  \bibinfo {address} {Cambridge, England},\ \bibinfo {year} {1992})\BibitemShut
  {NoStop}%
\bibitem [{\citenamefont {Robertson}(1934)}]{Robertson(34)_statUncePrinc}%
  \BibitemOpen
  \bibfield  {author} {\bibinfo {author} {\bibfnamefont {H.~P.}\ \bibnamefont
  {Robertson}},\ }\bibfield  {title} {\enquote {\bibinfo {title} {An
  indeterminacy relation for several observables and its classical
  interpretation},}\ }\href {\doibase 10.1103/PhysRev.46.794} {\bibfield
  {journal} {\bibinfo  {journal} {Phys. Rev.}\ }\textbf {\bibinfo {volume}
  {46}},\ \bibinfo {pages} {794--801} (\bibinfo {year} {1934})}\BibitemShut
  {NoStop}%
\bibitem [{\citenamefont {Heisenberg}(1927)}]{Heisenberg(27)_uc}%
  \BibitemOpen
  \bibfield  {author} {\bibinfo {author} {\bibfnamefont {W.}~\bibnamefont
  {Heisenberg}},\ }\bibfield  {title} {\enquote {\bibinfo {title} {{\"U}ber den
  anschaulichen inhalt der quantentheoretischen kinematik und mechanik},}\
  }\href {\doibase 10.1007/BF01397280} {\bibfield  {journal} {\bibinfo
  {journal} {Zeitschrift f{\"u}r Physik}\ }\textbf {\bibinfo {volume} {43}},\
  \bibinfo {pages} {172--198} (\bibinfo {year} {1927})}\BibitemShut {NoStop}%
\bibitem [{\citenamefont {Sudhir}\ \emph {et~al.}(2017)\citenamefont {Sudhir},
  \citenamefont {Wilson}, \citenamefont {Schilling}, \citenamefont {Sch\"utz},
  \citenamefont {Fedorov}, \citenamefont {Ghadimi}, \citenamefont
  {Nunnenkamp},\ and\ \citenamefont {Kippenberg}}]{Sudhir(16)_corr}%
  \BibitemOpen
  \bibfield  {author} {\bibinfo {author} {\bibfnamefont {V.}~\bibnamefont
  {Sudhir}}, \bibinfo {author} {\bibfnamefont {D.~J.}\ \bibnamefont {Wilson}},
  \bibinfo {author} {\bibfnamefont {R.}~\bibnamefont {Schilling}}, \bibinfo
  {author} {\bibfnamefont {H.}~\bibnamefont {Sch\"utz}}, \bibinfo {author}
  {\bibfnamefont {S.~A.}\ \bibnamefont {Fedorov}}, \bibinfo {author}
  {\bibfnamefont {A.~H.}\ \bibnamefont {Ghadimi}}, \bibinfo {author}
  {\bibfnamefont {A.}~\bibnamefont {Nunnenkamp}}, \ and\ \bibinfo {author}
  {\bibfnamefont {T.~J.}\ \bibnamefont {Kippenberg}},\ }\bibfield  {title}
  {\enquote {\bibinfo {title} {Appearance and disappearance of quantum
  correlations in measurement-based feedback control of a mechanical
  oscillator},}\ }\href {\doibase 10.1103/PhysRevX.7.011001} {\bibfield
  {journal} {\bibinfo  {journal} {Phys. Rev. X}\ }\textbf {\bibinfo {volume}
  {7}},\ \bibinfo {pages} {011001} (\bibinfo {year} {2017})}\BibitemShut
  {NoStop}%
\bibitem [{\citenamefont {Brooks}\ \emph {et~al.}(2012)\citenamefont {Brooks},
  \citenamefont {Botter}, \citenamefont {Schreppler}, \citenamefont {Purdy},
  \citenamefont {Brahms},\ and\ \citenamefont
  {Stamper-Kurn}}]{Brooks(12)_Squeez}%
  \BibitemOpen
  \bibfield  {author} {\bibinfo {author} {\bibfnamefont {Daniel W.~C.}\
  \bibnamefont {Brooks}}, \bibinfo {author} {\bibfnamefont {Thierry}\
  \bibnamefont {Botter}}, \bibinfo {author} {\bibfnamefont {Sydney}\
  \bibnamefont {Schreppler}}, \bibinfo {author} {\bibfnamefont {Thomas~P.}\
  \bibnamefont {Purdy}}, \bibinfo {author} {\bibfnamefont {Nathan}\
  \bibnamefont {Brahms}}, \ and\ \bibinfo {author} {\bibfnamefont {Dan~M.}\
  \bibnamefont {Stamper-Kurn}},\ }\bibfield  {title} {\enquote {\bibinfo
  {title} {Non-classical light generated by quantum-noise-driven cavity
  optomechanics},}\ }\href {\doibase 10.1038/nature11325} {\bibfield  {journal}
  {\bibinfo  {journal} {Nature}\ }\textbf {\bibinfo {volume} {488}},\ \bibinfo
  {pages} {476--480} (\bibinfo {year} {2012})}\BibitemShut {NoStop}%
\bibitem [{\citenamefont {Safavi-Naeini}\ \emph {et~al.}(2013)\citenamefont
  {Safavi-Naeini}, \citenamefont {Groblacher}, \citenamefont {Hill},
  \citenamefont {Chan}, \citenamefont {Aspelmeyer},\ and\ \citenamefont
  {Painter}}]{SafaviNaeini(13)_Squeezing}%
  \BibitemOpen
  \bibfield  {author} {\bibinfo {author} {\bibfnamefont {Amir~H.}\ \bibnamefont
  {Safavi-Naeini}}, \bibinfo {author} {\bibfnamefont {Simon}\ \bibnamefont
  {Groblacher}}, \bibinfo {author} {\bibfnamefont {Jeff~T.}\ \bibnamefont
  {Hill}}, \bibinfo {author} {\bibfnamefont {Jasper}\ \bibnamefont {Chan}},
  \bibinfo {author} {\bibfnamefont {Markus}\ \bibnamefont {Aspelmeyer}}, \ and\
  \bibinfo {author} {\bibfnamefont {Oskar}\ \bibnamefont {Painter}},\
  }\bibfield  {title} {\enquote {\bibinfo {title} {Squeezed light from a
  silicon micromechanical resonator},}\ }\href {\doibase 10.1038/nature12307}
  {\bibfield  {journal} {\bibinfo  {journal} {Nature}\ }\textbf {\bibinfo
  {volume} {500}},\ \bibinfo {pages} {185--189} (\bibinfo {year}
  {2013})}\BibitemShut {NoStop}%
\bibitem [{\citenamefont {Purdy}\ \emph {et~al.}(2013)\citenamefont {Purdy},
  \citenamefont {Yu}, \citenamefont {Peterson}, \citenamefont {Kampel},\ and\
  \citenamefont {Regal}}]{Purdy(13)_MechSqueeze}%
  \BibitemOpen
  \bibfield  {author} {\bibinfo {author} {\bibfnamefont {T.~P.}\ \bibnamefont
  {Purdy}}, \bibinfo {author} {\bibfnamefont {P.-L.}\ \bibnamefont {Yu}},
  \bibinfo {author} {\bibfnamefont {R.~W.}\ \bibnamefont {Peterson}}, \bibinfo
  {author} {\bibfnamefont {N.~S.}\ \bibnamefont {Kampel}}, \ and\ \bibinfo
  {author} {\bibfnamefont {C.~A.}\ \bibnamefont {Regal}},\ }\bibfield  {title}
  {\enquote {\bibinfo {title} {Strong optomechanical squeezing of light},}\
  }\href {\doibase 10.1103/PhysRevX.3.031012} {\bibfield  {journal} {\bibinfo
  {journal} {Phys. Rev. X}\ }\textbf {\bibinfo {volume} {3}},\ \bibinfo {pages}
  {031012} (\bibinfo {year} {2013})}\BibitemShut {NoStop}%
\bibitem [{\citenamefont {Buchmann}\ \emph {et~al.}(2016)\citenamefont
  {Buchmann}, \citenamefont {Schreppler}, \citenamefont {Kohler}, \citenamefont
  {Spethmann},\ and\ \citenamefont {Stamper-Kurn}}]{Buchmann(16)_Synodyne}%
  \BibitemOpen
  \bibfield  {author} {\bibinfo {author} {\bibfnamefont {L.~F.}\ \bibnamefont
  {Buchmann}}, \bibinfo {author} {\bibfnamefont {S.}~\bibnamefont
  {Schreppler}}, \bibinfo {author} {\bibfnamefont {J.}~\bibnamefont {Kohler}},
  \bibinfo {author} {\bibfnamefont {N.}~\bibnamefont {Spethmann}}, \ and\
  \bibinfo {author} {\bibfnamefont {D.~M.}\ \bibnamefont {Stamper-Kurn}},\
  }\bibfield  {title} {\enquote {\bibinfo {title} {Complex squeezing and force
  measurement beyond the standard quantum limit},}\ }\href {\doibase
  10.1103/PhysRevLett.117.030801} {\bibfield  {journal} {\bibinfo  {journal}
  {Phys. Rev. Lett.}\ }\textbf {\bibinfo {volume} {117}},\ \bibinfo {pages}
  {030801} (\bibinfo {year} {2016})}\BibitemShut {NoStop}%
\bibitem [{\citenamefont {Clark}\ \emph {et~al.}(2016)\citenamefont {Clark},
  \citenamefont {Lecocq}, \citenamefont {Simmonds}, \citenamefont {Aumentado},\
  and\ \citenamefont {Teufel}}]{Clark(16)_SqueezedLightRPSN}%
  \BibitemOpen
  \bibfield  {author} {\bibinfo {author} {\bibfnamefont {Jeremy~B.}\
  \bibnamefont {Clark}}, \bibinfo {author} {\bibfnamefont {Florent}\
  \bibnamefont {Lecocq}}, \bibinfo {author} {\bibfnamefont {Raymond~W.}\
  \bibnamefont {Simmonds}}, \bibinfo {author} {\bibfnamefont {Jos\'{e}}\
  \bibnamefont {Aumentado}}, \ and\ \bibinfo {author} {\bibfnamefont {John~D.}\
  \bibnamefont {Teufel}},\ }\bibfield  {title} {\enquote {\bibinfo {title}
  {Observation of strong radiation pressure forces from squeezed light on a
  mechanical oscillator},}\ }\href
  {http://www.nature.com/nphys/journal/vaop/ncurrent/full/nphys3701.html}
  {\bibfield  {journal} {\bibinfo  {journal} {Nat. Phys.}\ } (\bibinfo {year}
  {2016})}\BibitemShut {NoStop}%
\bibitem [{\citenamefont {Schnabel}\ \emph {et~al.}(2010)\citenamefont
  {Schnabel}, \citenamefont {Mavalvala}, \citenamefont {McClelland},\ and\
  \citenamefont {Lam}}]{Schnabel(10)_LIGOsque}%
  \BibitemOpen
  \bibfield  {author} {\bibinfo {author} {\bibfnamefont {Roman}\ \bibnamefont
  {Schnabel}}, \bibinfo {author} {\bibfnamefont {Nergis}\ \bibnamefont
  {Mavalvala}}, \bibinfo {author} {\bibfnamefont {David~E.}\ \bibnamefont
  {McClelland}}, \ and\ \bibinfo {author} {\bibfnamefont {Ping~K.}\
  \bibnamefont {Lam}},\ }\bibfield  {title} {\enquote {\bibinfo {title}
  {Quantum metrology for gravitational wave astronomy},}\ }\href {\doibase
  10.1038/ncomms1122} {\bibfield  {journal} {\bibinfo  {journal} {Nat.
  Communications}\ }\textbf {\bibinfo {volume} {1}},\ \bibinfo {pages} {1--10}
  (\bibinfo {year} {2010})}\BibitemShut {NoStop}%
\bibitem [{\citenamefont {Collaboration}(2011)}]{LIGO(11)_LIGOsque}%
  \BibitemOpen
  \bibfield  {author} {\bibinfo {author} {\bibfnamefont {The LIGO~Scientific}\
  \bibnamefont {Collaboration}},\ }\bibfield  {title} {\enquote {\bibinfo
  {title} {A gravitational wave observatory operating beyond the quantum
  shot-noise limit},}\ }\href {\doibase 10.1038/nphys2083} {\bibfield
  {journal} {\bibinfo  {journal} {Nat. Phys.}\ }\textbf {\bibinfo {volume}
  {7}},\ \bibinfo {pages} {962--965} (\bibinfo {year} {2011})}\BibitemShut
  {NoStop}%
\bibitem [{\citenamefont {Thompson}\ \emph {et~al.}(2008)\citenamefont
  {Thompson}, \citenamefont {Zwickl}, \citenamefont {Jayich}, \citenamefont
  {Marquardt}, \citenamefont {Girvin},\ and\ \citenamefont
  {Harris}}]{Thompson(08)_MemInMid}%
  \BibitemOpen
  \bibfield  {author} {\bibinfo {author} {\bibfnamefont {J.~D.}\ \bibnamefont
  {Thompson}}, \bibinfo {author} {\bibfnamefont {B.~M.}\ \bibnamefont
  {Zwickl}}, \bibinfo {author} {\bibfnamefont {A.~M.}\ \bibnamefont {Jayich}},
  \bibinfo {author} {\bibfnamefont {Florian}\ \bibnamefont {Marquardt}},
  \bibinfo {author} {\bibfnamefont {S.~M.}\ \bibnamefont {Girvin}}, \ and\
  \bibinfo {author} {\bibfnamefont {J.~G.~E.}\ \bibnamefont {Harris}},\
  }\bibfield  {title} {\enquote {\bibinfo {title} {Strong dispersive coupling
  of a high-finesse cavity to a micromechanical membrane},}\ }\href {\doibase
  10.1038/nature06715} {\bibfield  {journal} {\bibinfo  {journal} {Nature}\
  }\textbf {\bibinfo {volume} {452}},\ \bibinfo {pages} {72--75} (\bibinfo
  {year} {2008})}\BibitemShut {NoStop}%
\bibitem [{\citenamefont {Peterson}\ \emph {et~al.}(2016)\citenamefont
  {Peterson}, \citenamefont {Purdy}, \citenamefont {Kampel}, \citenamefont
  {Andrews}, \citenamefont {Yu}, \citenamefont {Lehnert},\ and\ \citenamefont
  {Regal}}]{Peterson(16)_BALcool}%
  \BibitemOpen
  \bibfield  {author} {\bibinfo {author} {\bibfnamefont {R.~W.}\ \bibnamefont
  {Peterson}}, \bibinfo {author} {\bibfnamefont {T.~P.}\ \bibnamefont {Purdy}},
  \bibinfo {author} {\bibfnamefont {N.~S.}\ \bibnamefont {Kampel}}, \bibinfo
  {author} {\bibfnamefont {R.~W.}\ \bibnamefont {Andrews}}, \bibinfo {author}
  {\bibfnamefont {P.-L.}\ \bibnamefont {Yu}}, \bibinfo {author} {\bibfnamefont
  {K.~W.}\ \bibnamefont {Lehnert}}, \ and\ \bibinfo {author} {\bibfnamefont
  {C.~A.}\ \bibnamefont {Regal}},\ }\bibfield  {title} {\enquote {\bibinfo
  {title} {Laser cooling of a micromechanical membrane to the quantum
  backaction limit},}\ }\href {\doibase 10.1103/PhysRevLett.116.063601}
  {\bibfield  {journal} {\bibinfo  {journal} {Phys. Rev. Lett.}\ }\textbf
  {\bibinfo {volume} {116}},\ \bibinfo {pages} {063601} (\bibinfo {year}
  {2016})}\BibitemShut {NoStop}%
\bibitem [{\citenamefont {Yu}\ \emph {et~al.}(2014)\citenamefont {Yu},
  \citenamefont {Cicak}, \citenamefont {Kampel}, \citenamefont {Tsaturyan},
  \citenamefont {Purdy}, \citenamefont {Simmonds},\ and\ \citenamefont
  {Regal}}]{Yu(14)_PnC}%
  \BibitemOpen
  \bibfield  {author} {\bibinfo {author} {\bibfnamefont {P.-L.}\ \bibnamefont
  {Yu}}, \bibinfo {author} {\bibfnamefont {K.}~\bibnamefont {Cicak}}, \bibinfo
  {author} {\bibfnamefont {N.~S.}\ \bibnamefont {Kampel}}, \bibinfo {author}
  {\bibfnamefont {Y.}~\bibnamefont {Tsaturyan}}, \bibinfo {author}
  {\bibfnamefont {T.~P.}\ \bibnamefont {Purdy}}, \bibinfo {author}
  {\bibfnamefont {R.~W.}\ \bibnamefont {Simmonds}}, \ and\ \bibinfo {author}
  {\bibfnamefont {C.~A.}\ \bibnamefont {Regal}},\ }\bibfield  {title} {\enquote
  {\bibinfo {title} {A phononic bandgap shield for high-q membrane
  microresonators},}\ }\href
  {http://scitation.aip.org/content/aip/journal/apl/104/2/10.1063/1.4862031}
  {\bibfield  {journal} {\bibinfo  {journal} {Appl. Phys. Lett.}\ }\textbf
  {\bibinfo {volume} {104}},\ \bibinfo {eid} {023510} (\bibinfo {year}
  {2014})}\BibitemShut {NoStop}%
\bibitem [{\citenamefont {Tsaturyan}\ \emph {et~al.}(2014)\citenamefont
  {Tsaturyan}, \citenamefont {Barg}, \citenamefont {Simonsen}, \citenamefont
  {Villanueva}, \citenamefont {Schmid}, \citenamefont {Schliesser},\ and\
  \citenamefont {Polzik}}]{Tsaturyan(14)_pnc}%
  \BibitemOpen
  \bibfield  {author} {\bibinfo {author} {\bibfnamefont {Y.}~\bibnamefont
  {Tsaturyan}}, \bibinfo {author} {\bibfnamefont {A.}~\bibnamefont {Barg}},
  \bibinfo {author} {\bibfnamefont {A.}~\bibnamefont {Simonsen}}, \bibinfo
  {author} {\bibfnamefont {L.~G.}\ \bibnamefont {Villanueva}}, \bibinfo
  {author} {\bibfnamefont {S.}~\bibnamefont {Schmid}}, \bibinfo {author}
  {\bibfnamefont {A.}~\bibnamefont {Schliesser}}, \ and\ \bibinfo {author}
  {\bibfnamefont {E.~S.}\ \bibnamefont {Polzik}},\ }\bibfield  {title}
  {\enquote {\bibinfo {title} {Demonstration of suppressed phonon tunneling
  losses in phononic bandgap shielded membrane resonators for high-q
  optomechanics},}\ }\href {\doibase 10.1364/OE.22.006810} {\bibfield
  {journal} {\bibinfo  {journal} {Opt. Express}\ }\textbf {\bibinfo {volume}
  {22}},\ \bibinfo {pages} {6810--6821} (\bibinfo {year} {2014})}\BibitemShut
  {NoStop}%
\bibitem [{\citenamefont {B\o{}rkje}\ \emph {et~al.}(2010)\citenamefont
  {B\o{}rkje}, \citenamefont {Nunnenkamp}, \citenamefont {Zwickl},
  \citenamefont {Yang}, \citenamefont {Harris},\ and\ \citenamefont
  {Girvin}}]{Boerkje(10)_Observa_RPSN_Theory}%
  \BibitemOpen
  \bibfield  {author} {\bibinfo {author} {\bibfnamefont {K.}~\bibnamefont
  {B\o{}rkje}}, \bibinfo {author} {\bibfnamefont {A.}~\bibnamefont
  {Nunnenkamp}}, \bibinfo {author} {\bibfnamefont {B.~M.}\ \bibnamefont
  {Zwickl}}, \bibinfo {author} {\bibfnamefont {C.}~\bibnamefont {Yang}},
  \bibinfo {author} {\bibfnamefont {J.~G.~E.}\ \bibnamefont {Harris}}, \ and\
  \bibinfo {author} {\bibfnamefont {S.~M.}\ \bibnamefont {Girvin}},\ }\bibfield
   {title} {\enquote {\bibinfo {title} {Observability of radiation-pressure
  shot noise in optomechanical systems},}\ }\href {\doibase
  10.1103/PhysRevA.82.013818} {\bibfield  {journal} {\bibinfo  {journal} {Phys.
  Rev. A}\ }\textbf {\bibinfo {volume} {82}},\ \bibinfo {pages} {013818}
  (\bibinfo {year} {2010})}\BibitemShut {NoStop}%
\bibitem [{\citenamefont {Aspelmeyer}\ \emph {et~al.}(2014)\citenamefont
  {Aspelmeyer}, \citenamefont {Kippenberg},\ and\ \citenamefont
  {Marquardt}}]{Aspelmeyer(14)_RevModPhys}%
  \BibitemOpen
  \bibfield  {author} {\bibinfo {author} {\bibfnamefont {Markus}\ \bibnamefont
  {Aspelmeyer}}, \bibinfo {author} {\bibfnamefont {Tobias~J.}\ \bibnamefont
  {Kippenberg}}, \ and\ \bibinfo {author} {\bibfnamefont {Florian}\
  \bibnamefont {Marquardt}},\ }\bibfield  {title} {\enquote {\bibinfo {title}
  {Cavity optomechanics},}\ }\href {\doibase 10.1103/RevModPhys.86.1391}
  {\bibfield  {journal} {\bibinfo  {journal} {Rev. Mod. Phys.}\ }\textbf
  {\bibinfo {volume} {86}},\ \bibinfo {pages} {1391--1452} (\bibinfo {year}
  {2014})}\BibitemShut {NoStop}%
\bibitem [{\citenamefont {Weinstein}\ \emph {et~al.}(2014)\citenamefont
  {Weinstein}, \citenamefont {Lei}, \citenamefont {Wollman}, \citenamefont
  {Suh}, \citenamefont {Metelmann}, \citenamefont {Clerk},\ and\ \citenamefont
  {Schwab}}]{Weinstein(14)_ObserInterpCool}%
  \BibitemOpen
  \bibfield  {author} {\bibinfo {author} {\bibfnamefont {A.~J.}\ \bibnamefont
  {Weinstein}}, \bibinfo {author} {\bibfnamefont {C.~U.}\ \bibnamefont {Lei}},
  \bibinfo {author} {\bibfnamefont {E.~E.}\ \bibnamefont {Wollman}}, \bibinfo
  {author} {\bibfnamefont {J.}~\bibnamefont {Suh}}, \bibinfo {author}
  {\bibfnamefont {A.}~\bibnamefont {Metelmann}}, \bibinfo {author}
  {\bibfnamefont {A.~A.}\ \bibnamefont {Clerk}}, \ and\ \bibinfo {author}
  {\bibfnamefont {K.~C.}\ \bibnamefont {Schwab}},\ }\bibfield  {title}
  {\enquote {\bibinfo {title} {Observation and interpretation of motional
  sideband asymmetry in a quantum electromechanical device},}\ }\href {\doibase
  10.1103/PhysRevX.4.041003} {\bibfield  {journal} {\bibinfo  {journal} {Phys.
  Rev. X}\ }\textbf {\bibinfo {volume} {4}},\ \bibinfo {pages} {041003}
  (\bibinfo {year} {2014})}\BibitemShut {NoStop}%
\bibitem [{\citenamefont {Botter}\ \emph {et~al.}(2012)\citenamefont {Botter},
  \citenamefont {Brooks}, \citenamefont {Brahms}, \citenamefont {Schreppler},\
  and\ \citenamefont {Stamper-Kurn}}]{Botter(12)_LinAmpModel}%
  \BibitemOpen
  \bibfield  {author} {\bibinfo {author} {\bibfnamefont {Thierry}\ \bibnamefont
  {Botter}}, \bibinfo {author} {\bibfnamefont {Daniel W.~C.}\ \bibnamefont
  {Brooks}}, \bibinfo {author} {\bibfnamefont {Nathan}\ \bibnamefont {Brahms}},
  \bibinfo {author} {\bibfnamefont {Sydney}\ \bibnamefont {Schreppler}}, \ and\
  \bibinfo {author} {\bibfnamefont {Dan~M.}\ \bibnamefont {Stamper-Kurn}},\
  }\bibfield  {title} {\enquote {\bibinfo {title} {Linear amplifier model for
  optomechanical systems},}\ }\href {\doibase 10.1103/PhysRevA.85.013812}
  {\bibfield  {journal} {\bibinfo  {journal} {Phys. Rev. A}\ }\textbf {\bibinfo
  {volume} {85}},\ \bibinfo {pages} {013812} (\bibinfo {year}
  {2012})}\BibitemShut {NoStop}%
\bibitem [{\citenamefont {Clerk}(2014)}]{ClerkLesHouches}%
  \BibitemOpen
  \bibfield  {author} {\bibinfo {author} {\bibfnamefont {Aashish~A.}\
  \bibnamefont {Clerk}},\ }\enquote {\bibinfo {title} {Quantum machines:
  Measurement and control of engineered quantum systems},}\ \ (\bibinfo
  {publisher} {Les Houche Summer School Lecture Notes, Oxford University
  Press},\ \bibinfo {year} {2014})\ Chap.\ \bibinfo {chapter} {Quantum noise
  and quantum measurement}\BibitemShut {NoStop}%
\end{thebibliography}

\begin{thebibliography}{7}%
\makeatletter
\providecommand \@ifxundefined [1]{%
 \@ifx{#1\undefined}
}%
\providecommand \@ifnum [1]{%
 \ifnum #1\expandafter \@firstoftwo
 \else \expandafter \@secondoftwo
 \fi
}%
\providecommand \@ifx [1]{%
 \ifx #1\expandafter \@firstoftwo
 \else \expandafter \@secondoftwo
 \fi
}%
\providecommand \natexlab [1]{#1}%
\providecommand \enquote  [1]{``#1''}%
\providecommand \bibnamefont  [1]{#1}%
\providecommand \bibfnamefont [1]{#1}%
\providecommand \citenamefont [1]{#1}%
\providecommand \href@noop [0]{\@secondoftwo}%
\providecommand \href [0]{\begingroup \@sanitize@url \@href}%
\providecommand \@href[1]{\@@startlink{#1}\@@href}%
\providecommand \@@href[1]{\endgroup#1\@@endlink}%
\providecommand \@sanitize@url [0]{\catcode `\\12\catcode `\$12\catcode
  `\&12\catcode `\#12\catcode `\^12\catcode `\_12\catcode `\%12\relax}%
\providecommand \@@startlink[1]{}%
\providecommand \@@endlink[0]{}%
\providecommand \url  [0]{\begingroup\@sanitize@url \@url }%
\providecommand \@url [1]{\endgroup\@href {#1}{\urlprefix }}%
\providecommand \urlprefix  [0]{URL }%
\providecommand \Eprint [0]{\href }%
\providecommand \doibase [0]{http://dx.doi.org/}%
\providecommand \selectlanguage [0]{\@gobble}%
\providecommand \bibinfo  [0]{\@secondoftwo}%
\providecommand \bibfield  [0]{\@secondoftwo}%
\providecommand \translation [1]{[#1]}%
\providecommand \BibitemOpen [0]{}%
\providecommand \bibitemStop [0]{}%
\providecommand \bibitemNoStop [0]{.\EOS\space}%
\providecommand \EOS [0]{\spacefactor3000\relax}%
\providecommand \BibitemShut  [1]{\csname bibitem#1\endcsname}%
\let\auto@bib@innerbib\@empty
\bibitem [{\citenamefont {Peterson}\ \emph {et~al.}(2016)\citenamefont
  {Peterson}, \citenamefont {Purdy}, \citenamefont {Kampel}, \citenamefont
  {Andrews}, \citenamefont {Yu}, \citenamefont {Lehnert},\ and\ \citenamefont
  {Regal}}]{Peterson(16)_BALcool}%
  \BibitemOpen
  \bibfield  {author} {\bibinfo {author} {\bibfnamefont {R.~W.}\ \bibnamefont
  {Peterson}}, \bibinfo {author} {\bibfnamefont {T.~P.}\ \bibnamefont {Purdy}},
  \bibinfo {author} {\bibfnamefont {N.~S.}\ \bibnamefont {Kampel}}, \bibinfo
  {author} {\bibfnamefont {R.~W.}\ \bibnamefont {Andrews}}, \bibinfo {author}
  {\bibfnamefont {P.-L.}\ \bibnamefont {Yu}}, \bibinfo {author} {\bibfnamefont
  {K.~W.}\ \bibnamefont {Lehnert}}, \ and\ \bibinfo {author} {\bibfnamefont
  {C.~A.}\ \bibnamefont {Regal}},\ }\bibfield  {title} {\enquote {\bibinfo
  {title} {Laser cooling of a micromechanical membrane to the quantum
  backaction limit},}\ }\href {\doibase 10.1103/PhysRevLett.116.063601}
  {\bibfield  {journal} {\bibinfo  {journal} {Phys. Rev. Lett.}\ }\textbf
  {\bibinfo {volume} {116}},\ \bibinfo {pages} {063601} (\bibinfo {year}
  {2016})}\BibitemShut {NoStop}%
\bibitem [{\citenamefont {Purdy}\ \emph {et~al.}(2013)\citenamefont {Purdy},
  \citenamefont {Yu}, \citenamefont {Peterson}, \citenamefont {Kampel},\ and\
  \citenamefont {Regal}}]{Purdy(13)_MechSqueezeSupp}%
  \BibitemOpen
  \bibfield  {author} {\bibinfo {author} {\bibfnamefont {T.~P.}\ \bibnamefont
  {Purdy}}, \bibinfo {author} {\bibfnamefont {P.-L.}\ \bibnamefont {Yu}},
  \bibinfo {author} {\bibfnamefont {R.~W.}\ \bibnamefont {Peterson}}, \bibinfo
  {author} {\bibfnamefont {N.~S.}\ \bibnamefont {Kampel}}, \ and\ \bibinfo
  {author} {\bibfnamefont {C.~A.}\ \bibnamefont {Regal}},\ }\bibfield  {title}
  {\enquote {\bibinfo {title} {Strong optomechanical squeezing of light},}\
  }\href {\doibase 10.1103/PhysRevX.3.031012} {\bibfield  {journal} {\bibinfo
  {journal} {Phys. Rev. X}\ }\textbf {\bibinfo {volume} {3}},\ \bibinfo {pages}
  {031012} (\bibinfo {year} {2013})}\BibitemShut {NoStop}%
\bibitem [{\citenamefont {Clerk}\ \emph {et~al.}(2010)\citenamefont {Clerk},
  \citenamefont {Devoret}, \citenamefont {Girvin}, \citenamefont {Marquardt},\
  and\ \citenamefont {Schoelkopf}}]{Clerk(10)_QNSupp}%
  \BibitemOpen
  \bibfield  {author} {\bibinfo {author} {\bibfnamefont {A.~A.}\ \bibnamefont
  {Clerk}}, \bibinfo {author} {\bibfnamefont {M.~H.}\ \bibnamefont {Devoret}},
  \bibinfo {author} {\bibfnamefont {S.~M.}\ \bibnamefont {Girvin}}, \bibinfo
  {author} {\bibfnamefont {Florian}\ \bibnamefont {Marquardt}}, \ and\ \bibinfo
  {author} {\bibfnamefont {R.~J.}\ \bibnamefont {Schoelkopf}},\ }\bibfield
  {title} {\enquote {\bibinfo {title} {Introduction to quantum noise,
  measurement, and amplification},}\ }\href {\doibase
  10.1103/RevModPhys.82.1155} {\bibfield  {journal} {\bibinfo  {journal} {Rev.
  Mod. Phys.}\ }\textbf {\bibinfo {volume} {82}},\ \bibinfo {pages}
  {1155--1208} (\bibinfo {year} {2010})}\BibitemShut {NoStop}%
\bibitem [{\citenamefont {B\o{}rkje}\ \emph {et~al.}(2010)\citenamefont
  {B\o{}rkje}, \citenamefont {Nunnenkamp}, \citenamefont {Zwickl},
  \citenamefont {Yang}, \citenamefont {Harris},\ and\ \citenamefont
  {Girvin}}]{Boerkje(10)_Observa_RPSN_TheorySupp}%
  \BibitemOpen
  \bibfield  {author} {\bibinfo {author} {\bibfnamefont {K.}~\bibnamefont
  {B\o{}rkje}}, \bibinfo {author} {\bibfnamefont {A.}~\bibnamefont
  {Nunnenkamp}}, \bibinfo {author} {\bibfnamefont {B.~M.}\ \bibnamefont
  {Zwickl}}, \bibinfo {author} {\bibfnamefont {C.}~\bibnamefont {Yang}},
  \bibinfo {author} {\bibfnamefont {J.~G.~E.}\ \bibnamefont {Harris}}, \ and\
  \bibinfo {author} {\bibfnamefont {S.~M.}\ \bibnamefont {Girvin}},\ }\bibfield
   {title} {\enquote {\bibinfo {title} {Observability of radiation-pressure
  shot noise in optomechanical systems},}\ }\href {\doibase
  10.1103/PhysRevA.82.013818} {\bibfield  {journal} {\bibinfo  {journal} {Phys.
  Rev. A}\ }\textbf {\bibinfo {volume} {82}},\ \bibinfo {pages} {013818}
  (\bibinfo {year} {2010})}\BibitemShut {NoStop}%
\bibitem [{\citenamefont {Buchmann}\ \emph {et~al.}(2016)\citenamefont
  {Buchmann}, \citenamefont {Schreppler}, \citenamefont {Kohler}, \citenamefont
  {Spethmann},\ and\ \citenamefont
  {Stamper-Kurn}}]{Buchmann(16)_Synodyne_Supp}%
  \BibitemOpen
  \bibfield  {author} {\bibinfo {author} {\bibfnamefont {L.~F.}\ \bibnamefont
  {Buchmann}}, \bibinfo {author} {\bibfnamefont {S.}~\bibnamefont
  {Schreppler}}, \bibinfo {author} {\bibfnamefont {J.}~\bibnamefont {Kohler}},
  \bibinfo {author} {\bibfnamefont {N.}~\bibnamefont {Spethmann}}, \ and\
  \bibinfo {author} {\bibfnamefont {D.~M.}\ \bibnamefont {Stamper-Kurn}},\
  }\bibfield  {title} {\enquote {\bibinfo {title} {Complex squeezing and force
  measurement beyond the standard quantum limit},}\ }\href {\doibase
  10.1103/PhysRevLett.117.030801} {\bibfield  {journal} {\bibinfo  {journal}
  {Phys. Rev. Lett.}\ }\textbf {\bibinfo {volume} {117}},\ \bibinfo {pages}
  {030801} (\bibinfo {year} {2016})}\BibitemShut {NoStop}%
\bibitem [{\citenamefont {Clerk}\ \emph {et~al.}(2008)\citenamefont {Clerk},
  \citenamefont {Marquardt},\ and\ \citenamefont
  {Jacobs}}]{Clerk(08)_BAEsqueezingSupp}%
  \BibitemOpen
  \bibfield  {author} {\bibinfo {author} {\bibfnamefont {A.~A.}\ \bibnamefont
  {Clerk}}, \bibinfo {author} {\bibfnamefont {F.}~\bibnamefont {Marquardt}}, \
  and\ \bibinfo {author} {\bibfnamefont {K.}~\bibnamefont {Jacobs}},\
  }\bibfield  {title} {\enquote {\bibinfo {title} {Back-action evasion and
  squeezing of a mechanical resonator using a cavity detector},}\ }\href
  {http://stacks.iop.org/1367-2630/10/i=9/a=095010} {\bibfield  {journal}
  {\bibinfo  {journal} {New J. Phys.}\ }\textbf {\bibinfo {volume} {10}},\
  \bibinfo {pages} {095010} (\bibinfo {year} {2008})}\BibitemShut {NoStop}%
\bibitem [{\citenamefont {Braginsky}\ \emph {et~al.}(1980)\citenamefont
  {Braginsky}, \citenamefont {Vorontsov},\ and\ \citenamefont
  {Thorne}}]{Braginsky(80)_QNDdefSupp}%
  \BibitemOpen
  \bibfield  {author} {\bibinfo {author} {\bibfnamefont {Vladimir~B.}\
  \bibnamefont {Braginsky}}, \bibinfo {author} {\bibfnamefont {Yuri~I.}\
  \bibnamefont {Vorontsov}}, \ and\ \bibinfo {author} {\bibfnamefont {Kip~S.}\
  \bibnamefont {Thorne}},\ }\bibfield  {title} {\enquote {\bibinfo {title}
  {Quantum nondemolition measurements},}\ }\href {\doibase
  10.1126/science.209.4456.547} {\bibfield  {journal} {\bibinfo  {journal}
  {Science}\ }\textbf {\bibinfo {volume} {209}},\ \bibinfo {pages} {547--557}
  (\bibinfo {year} {1980})}\BibitemShut {NoStop}%
\end{thebibliography}

%

\end{document}